\documentclass{article}

\usepackage{arxiv}
\usepackage{longtable}

\usepackage[utf8]{inputenc} 
\usepackage[T1]{fontenc}    
\usepackage{hyperref}       
\usepackage{url}            
\usepackage{booktabs}       
\usepackage{amsfonts}       
\usepackage{nicefrac}       
\usepackage{microtype}      
\usepackage{lipsum}
\usepackage{graphicx}
\graphicspath{ {./images/} }

\usepackage{amsfonts}   
\usepackage{amssymb}    
\usepackage{amsmath}
\usepackage[table]{xcolor}

\usepackage{cleveref}
\crefrangelabelformat{equation}{(#3#1#4--#5#2#6)}
\crefname{equation}{Eq.}{Eqs.}
\Crefname{equation}{Equation}{Equations}

\usepackage{amsthm}
\usepackage{mathtools}
\usepackage{amsfonts} 

\usepackage{arxiv}
\usepackage{algorithm, algorithmic}

\usepackage[nocomma]{optidef}

\usepackage{empheq}
\usepackage{graphicx}

\usepackage[utf8]{inputenc} 
\usepackage[T1]{fontenc}    
\usepackage{hyperref}       
\usepackage{url}            
\usepackage{amsfonts}       
\usepackage{booktabs}       
\usepackage{nicefrac}       
\usepackage{microtype}      
\usepackage{lipsum}
\usepackage{graphicx}
\graphicspath{ {./images/} }
\usepackage{float}
\usepackage{caption}
\usepackage[font=small,labelfont=bf, labelsep=endash, labelfont=bf, justification=raggedright, width=1.33\linewidth]{caption} 

\usepackage{enumitem}
\setlist[enumerate]{itemsep=10pt, topsep=10pt, parsep=0pt}

\usepackage[utf8]{inputenc} 
\usepackage[T1]{fontenc}    
\usepackage{hyperref}       
\usepackage{url}            
\usepackage{booktabs}       
\usepackage{amsfonts}       
\usepackage{nicefrac}       
\usepackage{microtype}      
\usepackage{lipsum}
\usepackage{graphicx}
\graphicspath{ {./images/} }

\usepackage{csquotes}

\numberwithin{equation}{section}

\DeclareMathSymbol{\sm}{\mathbin}{AMSa}{"39}

\theoremstyle{definition}
\newtheorem{defn}{Definition}[section]

\newtheorem{assumption}{Assumption}[section]

\newcommand{\dt}{\,\mathrm{dt}}

\newcommand{\du}{\,\mathrm{du}}

\newcommand{\pmat}[1]{\begin{pmatrix}#1\end{pmatrix}}

\newcommand{\smallsec}[1]{\smallskip\textit{#1}\smallskip}

\renewcommand{\S}{\mathcal{S}}
\newcommand{\T}{\mathcal {T}}

\newcommand{\R}{\mathbf{R}} 

\newcommand{\smi}{{s_{\! \sm i}}}

\newcommand{\types}{\{t_1^{k_1}, \dots, t_n^{k_n}\}}

\newcommand{\hatsmi}{{\hat s}_{\sm i}}
\newcommand{\mi}{{\sm i}}

\newcommand{\set}[1]{\left\{#1\right\}}

\usepackage{authblk}

\title{Position building in competition is a game with incomplete information\thanks{This is version two. The changes are corrections of numerous typos and inconsistencies between data in tables and plots.}}

\author{
  Neil A. Chriss\\
  \texttt{neil.chriss@gmail.com} \\
}

\begin{document}
\maketitle
\begin{abstract}
This paper examines strategic trading under incomplete information, where firms lack full knowledge of key aspects of their competitors' trading strategies such as target sizes and market impact models.  We extend previous work on competitive trading equilibria by incorporating uncertainty through the framework of Bayesian games. This allows us to analyze scenarios where firms have diverse beliefs about market conditions and each other's strategies.  We derive optimal trading strategies in this setting and demonstrate how uncertainty significantly impacts these strategies compared to the complete information case.

Furthermore, we introduce a novel approach to model the presence of non-strategic traders, even when strategic firms disagree on their characteristics.  Our analysis reveals the complex interplay of beliefs and strategic adjustments required in such an environment.  Finally, we discuss limitations of the current model, including the reliance on linear market impact and the lack of dynamic strategy adjustments, outlining directions for future research.
\end{abstract}


\section{Introduction}
\label{sec:introduction}


In some circumstances, trading is a \textit{strategic} interaction between market participants who are trading the same asset over the same period of time. This is the case where firms are concerned with the collective impact of their trades on market prices and choose how and when to trade in order to minimize their overall cost of trading, taking into consideration what the other firms are likely to do. The general situation just described, which we call \textit{ trading in competition}, was studied in \cite{chriss2024optposnbldg} and \cite{chriss2024equilibria} where it was shown that there unique Nash equilibria always exist (in the sense of \cite{nash_1950}, \cite{nash_1951} and Definition \ref{def:nash-equilibrium} below) that only depend on the number of firms involved, their target quantities, and the market impact model the firms use to select their trades. 

A key assumption and in our view a major shortcoming of our previous work is that it implicitly assumes that trading in competition is a game with \textit{complete information}, one for which participants are in possession of full knowledge of all relevant aspects of the game, which in this context means the number of firms, their target quantities and the market impact model they use. By contrast, in most real-world situations, trading in competition is a game with \textit{incomplete information}, in which some or all participants lack full knowledge of at least some aspects of the game. In the case of trading, firms do not know their competitors' target quantities, and are far from sure of precise values of the market impact parametersor even \textit{which} models they use.

The distinction between games with complete and incomplete information can be traced back \textit{at least} to von Neumann and Morgenstern's foundational work on game theory \cite{von2007theory}. Early in their discussion, they state that "... we cannot avoid the assumption that all subjects of the economy under consideration are completely informed about the physical characteristics of the situation in which they operate..." They refer to this as "complete information" and note:

\begin{quote}
    The nature and importance of this assumption has been given extensive attention in the literature and the subject is probably very far from being exhausted. We propose not to enter upon it. The question is too vast and too difficult and we believe that it is best to "divide difficulties." I.e. we wish to avoid this complication which, while interesting in its own right, should be considered separately from our present problem.
\end{quote}

For more details on von Neumann and Morgenstern's treatment of incomplete information, see \cite{myerson2004comments}. The main point of this discussion is that incomplete information is a \textit{persistent} feature in trading situations due to institutional, regulatory and competitive factors. 

The present paper aims to demonstrate how uncertainty can be incorporated into competitive trading using the framework of games with "Bayesian" players (or Bayesian games), as developed in John Harsanyi's seminal work on games with incomplete information \cite{harsanyi1967games,harsanyi1994game}. Harsanyi’s contributions provide a general framework for addressing \textit{information uncertainty} in game theory and extend Nash's equilibrium concept.

Building on this foundation, we extend the work on trading in competition presented in \cite{chriss2024equilibria} and \cite{chriss2024optposnbldg} by allowing for incomplete information. To achieve this, we first review the theory of games with Bayesian players in Section \ref{sec:games-incomplete}, where we detail the methods Harsanyi introduced to handle incomplete information. We then apply these concepts to the context of competitive trading. 

In Section \ref{sec:trading-and-game-theory}, we revisit the fundamental game-theoretic concepts employed in \cite{chriss2024equilibria} and \cite{chriss2024optposnbldg}, providing the necessary groundwork for our analysis. Section \ref{sec:games-incomplete} introduces the framework of Harsanyi's Bayesian games, setting the stage for an in-depth exploration of incomplete information within the trading context. A central result of this approach is a method for transforming a game with incomplete information into an equivalent game of complete information. This transformation leverages Harsanyi's framework to ensure that the resulting game retains the essential features of the original, but can be analyzed as a complete information game.

The advantage of this transformation is significant: complete information games are guaranteed to have at least one Nash equilibrium\footnote{It is a fundamental result that all $n$-player non-cooperative complete information games have at least one Nash equilibrium \cite{nash_1951}.}. The equilibria derived in the transformed game can then be mapped back to the original incomplete information game, yielding strategies that are not only viable, but also exhibit properties interpretable as optimal within the context of competitive trading.

\smallsec{Basic outline of the paper}

The main result of this paper is a description of $n$-firm equilibria in trading in games with incomplete information, building on the framework introduced in \cite{chriss2024equilibria}. This framework is based on the following assumptions, which are reviewed in detail in Section \ref{sec:market-impact-models}:

\begin{enumerate}
    \item There are $n$ firms that trade \textit{ in competition with each other} over a fixed and common period of time (the "trading period").
    
    \item Each firm has a target quantity of stock it wishes to acquire before the end of the trading period.
    \item At the outset, each firm chooses a trading strategy that provides an ex-ante complete description of how it will trade during the trading period.
    
    \item Each firm uses the market impact model from the Almgren-Chriss framework (see \cite{almgren2001optimal}), a linear model with both permanent and temporary impact\footnote{The model is linear in the sense that the impact on prices is linear in the quantity traded and therefore the cost of trading \textit{quadratic} is quadratic in quantity traded.}. This model features a single market impact parameter, referred to as the \textit{market impact coefficient}.
\end{enumerate}

The novelty of this paper lies in relaxing a key assumption implicit in \cite{chriss2024equilibria}, namely that each firm's target quantity and market impact parameter are \textit{common knowledge}. In game theory, a fact is considered common knowledge among players if (i) every player knows the fact, (ii) every player knows that every other player knows the fact, (iii) every player knows that every other player knows that every player knows the fact, and so on. The concept of common knowledge was explicitly introduced to game theory only in the 1970s\footnote{The term "common knowledge" may have originated in David Lewis' philosophical treatise \cite{lewis2008convention} and was formally defined in game theory by Aumann in his seminal work \cite{aumann1976agree}, though it can be argued that the idea was \textit{implicitly} present in earlier works, such as those of von Neumann and Morgenstern.}.

In Section \ref{sec:games-incomplete}, we provide a self-contained review of Bayesian games, following Harsanyi's approach as outlined in his Nobel Prize lecture \cite{harsanyi1994game}\footnote{In this lecture, Harsanyi presents his "type-centric" approach to Bayesian games, which is conceptually equivalent to but more accessible than his original formulation in \cite{harsanyi1967games}.}. Readers familiar with this theory may safely skip this section. Section \ref{sec:trading-and-game-theory} reviews the trading and market setup necessary for the subsequent analysis. In particular, Section \ref{sec:solving-bayesian} formulates trading in competition under incomplete information. We present the two-firm case in full detail and provide an outline of the $n$-firm results, as they follow similar reasoning but involve more complex notation.

\smallsec{What is new?}

The approach of treating trading in competition as a game with incomplete information appears to be novel. In Section \ref{sec:two-firm-examples-1}, we demonstrate that uncertainty about what firms do and do not know regarding other firms' target quantities and beliefs about market impact parameters can have significant—and sometimes profound—effects on optimal strategies. These effects stand in contrast to the strategies that would emerge in the absence of such uncertainty.

A new result presented in this paper concerns the incorporation of \textit{non-strategic} firms into the model. Non-strategic firms are market participants that trade without considering the actions of other firms—for instance, when they believe they are so large relative to their competition that their primary concern is to spread their trades as evenly as possible over the trading period. In Section \ref{sec:non-strategic-firm}, we show how to model scenarios involving non-strategic firms, even in cases where there is disagreement among strategic firms about key aspects of the non-strategic firm, such as the size of its trades or the likelihood of its participation. Section \ref{sec:non-strategic-example} provides examples to illustrate these dynamics. In one such example, we analyze two firms (labeled 1 and 2) and show that even if firm 1 does not believe a non-strategic firm will participate, firm 1 must still adjust its strategy if there is any chance that firm 2 believes firm 1 might think the non-strategic firm \textit{will} participate. This interplay highlights the intricate adjustments required when dealing with incomplete information.

\smallsec{What is left to be done?}

While this paper demonstrates the importance of viewing trading in the context of games with incomplete information, there remain significant areas for further exploration on the technical side. The most pressing issues are as follows:

\begin{enumerate}
    \item \textit{Market impact models:} This paper relies on linear market impact models, as used in \cite{almgren2001optimal} and \cite{chriss2024equilibria}, among others. While we allow for uncertainty regarding certain \textit{parameters} of the model (specifically, the market impact parameter), we assume it is common knowledge that all firms use this fundamental linear framework.

    \item \textit{Lack of dynamics:} As noted earlier, this paper and its predecessors, including \cite{chriss2024equilibria}, derive equilibria in terms of \textit{ex ante} trading strategies. The setup treats trading as a game where "moves" are fixed trading strategies determined prior to the start of trading and remain unchanged throughout. In real-world scenarios, however, information revealed during the trading period would naturally influence trading decisions as events unfold.
\end{enumerate}

Regarding point 1, it is, in \textit{principle}, possible to extend the ideas in this paper to other market impact models\footnote{An extensive literature exists on this subject; see references in \cite{chriss2024equilibria}, \cite{chriss2024optposnbldg}, and, for example, \cite{Gatheral2010three}.}. Additionally, incorporating uncertainty about which market impact model a firm might use could be approached similarly to how Section \ref{sec:non-strategic-firm} models uncertainty regarding non-strategic firms. The main challenge in extending this aspect lies less in the \textit{mathematics} and more in the \textit{computational} complexity, an issue we leave open for future work.

As for point 2, the strategies derived here represent the actions firms \textit{would adopt} if they and all other firms were committed to their initial strategies and unable to adjust them during the trading period. The validity of this assumption hinges on two factors: (i) the likelihood of deviations from predetermined strategies during trading, and (ii) the impact that the possibility of such deviations would have on the strategies firms choose at the outset. These factors depend on the length of the trading period, the volatility of the underlying markets, and the nature of the market participants themselves.

\section{Games with incomplete information}
\label{sec:games-incomplete}

This section provides essential background on games with incomplete information and games with "Bayesian" players, drawing on John Harsanyi's seminal work on modeling information uncertainty in games (see \cite{harsanyi1967games} and \cite{harsanyi1994game}). In order to properly address these concepts, we present a self-contained account of the theory, preparing for its application to trading in competition, which is developed in Section \ref{sec:trading-in-comp-bayesian}. In this section, we primarily discuss the concepts in terms of "games" and "players," transitioning to "trading" and "firms" in later sections.

We believe that assuming \textit{complete information} when it comes to trading in competition warrants skepticism. It is unrealistic to assume that every firm knows how much every other firm intends to trade or their assumptions about market impact parameters. Similarly, it seems implausible that every firm is fully aware of what other firms believe and intend regarding their trading strategies. Even if all of this information were somehow known, questions remain about \textit{why} firms should be expected to adopt equilibrium strategies—especially when these strategies might counter intuition or when there is little to no penalty for deviating from them. As Aumann notes in \cite{aumann1987correlated}, "Nash equilibrium does make sense if one starts by assuming that, for some specified reason, each player knows which strategies the other players are using. But this assumption appears rather restrictive."

\subsection{Game theoretic background}
\label{sec:game-theory-defns}

The overarching framework for modeling strategic interactions among players is the $n$-player non-cooperative game, whose central solution concept is the celebrated Nash equilibrium (see \cite{nash_1951}). For completeness, we provide an overview of both, beginning with the formal definition of an $n$-player non-cooperative game.

\medskip

\begin{defn}[Non-cooperative game]
\label{def:non-cooperative-game}
An $n$-player non-cooperative game consists of the $n$ players, identified simply as $i=1, \dots, n$. Each player is characterized by the following:

\begin{enumerate}
    \item \textit{Pure strategies:} Each player has a set of \textit{pure strategies} or \textit{moves} available to them. These represent the specific actions or "plays" that a player can make. For example, in the two-player game of rock-paper-scissors, each player has three admissible actions: "Rock," "Paper," and "Scissors." For player $i$, an action is denoted by $a_{ij}$, where $j$ indexes the possible actions available to that player.

    \item \textit{Mixed strategies:} A \textit{mixed strategy} for player $i$ is a weighted sum of pure strategies, expressed as $\sum_j \lambda_j a_{ij}$, where $\lambda_j \geq 0$ for all $j$, and $\sum_j \lambda_j = 1$. Mixed strategies can be interpreted as the outcomes of a random process, with $\lambda_j$ representing the probability that player $i$ selects action $a_{ij}$.

    \item \textit{Payoff functions:} Each player $i$ has a \textit{payoff function}, $V_i$, which maps a strategy $n$-tuple to a numerical value representing that player's payoff. Payoff functions capture the value or utility a player associates with a given outcome. Importantly, the notion of \textit{value} here is subjective, reflecting the utility of an outcome from the perspective of the individual player.
\end{enumerate}

A $n$-player game is \textit{non-cooperative} when there is no communication or coordination between the players.
\end{defn}

A natural question that arises in the study of $n$-player non-cooperative games is whether there exist particular $n$-tuples of strategies that possess special properties. Nash, in his seminal work \cite{nash_1951}, addressed this question by introducing the concept of the Nash equilibrium. We define it as follows: Let $\hatsmi = (s_1, \dots, s_{i-1}, \hat{s}_i, s_{i+1}, \dots, s_n)$ denote a strategy profile obtained by modifying $s$ to replace $s_i$ with $\hat{s}_i \in S_i$, representing a deviation by player $i$.

\medskip

\begin{defn}[Nash equilibrium]
\label{def:nash-equilibrium}
Given an $n$-player non-cooperative game with strategy sets $S_i$ for players $i=1, \dots, n$, and payoff functions $V_i$ a Nash equilibrium is a strategy $n$-tuple $s=(s_1, \dots, s_n)$ if for any player $i$ and any $\hatsmi$ that replaces $s_i$ with $\hat s_i$ in $s$ for $\hat s_i \in S_i$ we have

\begin{equation}
    V_i(\hatsmi) \le V_i(s) 
\end{equation}

In other words, if all players are player their components of a Nash equilibrium strategy $s$, then any player who chooses to deviate from this equilibrium playing $\hat s_i$ instead of $s_i$, will not gain from such a move. Put differently, no player has any incentive to deviate from a Nash equilibrium strategy.
\end{defn}

Nash proved in \cite{nash_1951} that any $n$-player non-cooperative game there is at least one Nash equilibrium. It is interesting to and perhaps surprising that \cite{nash_1951}, though mathematically elegant and precise, never explicitly mentions the \textit{implicit assumption}, that in the games he studies, the available pure strategies and payoff functions associated with each player are \textit{common knowledge}, meaning that every player knows what every other player's pure strategies and payoff functions are and moreover that every player knows that every other player knows these as well. Such a game is known as a \textit{complete information} game. In \cite{chriss2024equilibria} we showed that $n$-traders "compete" for liquidity in a stock then for a given set of payoff functions and target quantities there is a unique Nash equilibrium set of strategies. As in \cite{nash_1951}, \cite{chriss2024equilibria} never explicitly mentions that the target quantities and payoff functions are common knowledge.  

The next sections describe the transition from \textit{complete information games} to games with \textit{incomplete information} in which it is no longer assumed that basic features of the game are common knowledge to all of the players. 

\subsection{Knowledge, information and beliefs}
\label{sec:types-1}

To start, we review what games with incomplete information are, following closely Harsanyi's seminal work in \cite{harsanyi1967games} and \cite{harsanyi1994game}. To clarify matters at the outset there is a critical distinction between games with incomplete information on the one hand, and \textit{Bayesian games}, or games with Bayesian players, on the other. Games with incomplete information are those that belong to the broad category of games in which some or all of the players lack full knowledge of critical aspects of the game or the players. Bayesian games, on the other hand, are a theoretical construct that models uncertainty in games by imagining that each participant in the game has one or more \textit{hypothetical} types that capture all of the relevant information about the moves they can play in the game, their beliefs about other players and their payoff functions. We start by defining games with incomplete information noting that as we will soon see a strictly formal definition is difficult to come by. 

\medskip

\begin{defn}[Incomplete information game (informal)]
\label{def:incomplete-information-game}
An $n$-player game of incomplete information is a game in which one or more of the players lack full information about aspects of the game that impact how players select their strategies. For example, they make lack full information about other players' payoff functions, available moves, intentions or beliefs concerning the other players. 
\end{defn}

\medskip

\textit{Discussion}

We start with an unspecified $n$-player non-cooperative game in which players $i=1, \dots, n$ simultaneously play an $n$-tuple of strategies $s=(s_1, \dots, s_n)$. Then there are three general categories of information in games with incomplete information (see \cite{harsanyi1967games}, Part I):

\begin{enumerate}
    \item Players may not know the "physical outcome function" of the game. That is, for every strategy $n$-tuple the players can play, there is a "ground truth" physical reality that happens and is common to all players. For example, in trading, such an outcome function is the reaction of market prices to the aggregate trading of all the firms involved. 

    \item Players may not have full knowledge of their own or (more likely) other players \textit{payoff functions}. The idea here is that since players select their strategies not on outcomes but on how they \textit{value} outcomes, this sort of knowledge impacts their strategy selection.

    \item Players may lack full information about their own or (more likely) other players available strategies. In trading, individual firms may face restrictions on how they can trade, due to, for example, capital or short-selling constraints.
\end{enumerate}

As Harsanyi points out, \textit{all} cases of incomplete information may be reduced to one of the above three. These three cases do even more than is at first apparent because they also capture what one player (say, player $i$) knows or believes about another player (say, $j$) with respect to one or more of the above considerations. In general, these considerations represent a complete description of all the ways in which a game can be incomplete and it's interesting to touch upon the context in which this originally arose, which in fact was the very real-world problem of nuclear disarmament negotiations. 

From 1965 to 1969 the U.S. Arms Control and Disarmament Agency employed a group of game-theorists as consultants\footnote{The group assembled was a veritable who's who of luminaries, including no less than four Nobel laureates: John Harsanyi, Reinhardt Selten and Robert Aumann and Gerard Debreu.}. The problem was that both the U.S. and the (then) Soviet Union wanted to slow down the growth of their nuclear weapons stockpiles, but to do so required assurances that the other side would follow suit. At the heart of their concerns was uncertainty concerning information. As Harsanyi puts it in \cite{harsanyi1994game}:

    \begin{quote}
    I realized that a major problem in arms control negotiations is the fact that each side is relatively \textit{well informed} about \textit{its own position} with respect to various variables relevant to arms control negotiations, such as its own policy objectives, its peaceful or bellicose attitudes toward the other side, its military strength, its own ability to introduce new military technologies, and so on -- but may be \textit{rather poorly informed} about the \textit{other side's} position in terms of such variables.    
    \end{quote}
    
Thus Harsanyi formulated his approach to games with incomplete information which we take up now. To begin, we note Harsanyi's seminal work on the subject first appeared in \cite{harsanyi1967games}, but it is acknowledged that this formulation has some unnecessary complications (see, for example, \cite{myerson2004comments}) and we follow Harsanyi's own description in his Nobel prize lecture (see \cite{harsanyi1994game}) which is equivalent but easier to follow. 

\medskip

\textit{Incomplete information as an intractable problem}

\medskip

At first blush it may not be be clear whether games with incomplete information are an easy, difficult or \textit{intractable} problem. In \cite{harsanyi1967games} Part I, Section 1, Harsanyi demonstrates that incomplete information in many circumstances spins up an infinite hierarchy of beliefs which is, for all intents and purposes, intractable. Consider a single parameter $X$ of the game for which some player $i$ is uncertain of its value. Then not only does player $i$ need to consider their own beliefs about this parameter but must also the beliefs of other players. 

According to Bayesian theory, player $i$'s beliefs concerning the parameter may be summarized by a probability distribution $e_i X$ over the possibly values $X$. We call this player $i$'s \textit{first-order beliefs} about the parameter $X$. Clearly $e_i X$ is important in understanding how player $i$ will go about selecting its strategies, but clearly since this is so, then in order for a different player $j$ to select its strategies it must take into account how player $i$ will select its strategies which in turn requires the to form beliefs about player $i$'s beliefs, which we denote $e_j\,e_i\,X$ and refer to as its \textit{second-order beliefs}. Given this it is clear that for yet a different player $k$ to contemplate player $j$'s and player $i$'s strategies it must form views $e_k\,e_j\, e_i X$ and $e_k\, e_i K$, which are a form of \textit{third-order beliefs}. Similarly player $i$ must form its own third-order beliefs $e_i e_j e_i X$ to understand how player $j$'s beliefs about its beliefs about $X$. Fourth-order and higher-order beliefs may be defined analogously. The upshot is that a single parameter about which a single player is uncertain spawns an infinite hierarchy of beliefs which must be taken into consideration in strategy selection among all players. From this perspective it is easy to countenance von Neumann and Morgenstern's the question is "too vast" comment (see Section \ref{sec:introduction}) concerning incomplete information. Therefore one may ask what alternatives are available, and this is where Harsany's games with "Bayesian" players make an entrance. In order to get there we must take a short detour to discuss payoff functions.

As a general matter one of the key contributions of \cite{harsanyi1967games} is that all sources of uncertainty may be reduced to the uncertainty concerning the payoff functions of each. The value of this is that the entirety of the problem can be modeled as one in which firms must workout what the payoff functions of the other firms are. We briefly sketch how various forms of uncertainty reduce to uncertainty concerning payoff functions here.

To begin, consider case 3, uncertainty about strategy spaces. Following \cite{harsanyi1967games} Part I, Section 2, we see that if firm $i$ is not allowed to trade a given strategy $s_i^0$ then this is equivalent to the assumption that firm $i$ will simply never trade this strategy, which in turn is equivalent to firm $i$'s payoff in any case where it trades $s_i$ is extremely negative. That is, we assume that for any set of strategies $(s_1, \dots, s_i^0, \dots, s_n)$ the payoff to firm $i$, $V_i(s_1, \dots, s_i^0, \dots, s_n)$ is extremely negative . 

Cases 1 and 2 can also be reduced to uncertainty concerning payoff functions because market impact functions in the case of trading in competition go directly into the payoff functions of each firm as laid out in Section \ref{sec:mathematical-formulation} and Definition \ref{def:imp-cost}. A basic premise in the case of trading is that each firm seeks to minimize its implementation cost and measures their payoff as their expected gain from owning the stock less the implementation cost. Thus the  market impact model is a component of the payoff function itself. In an exactly similar way uncertainty about target quantities is a direct component of the payoff function.

\subsection{Bayesian players and player types}
\label{sec:types}
\label{sec:bayesian-equilibrium}

A high-level outline of Harsanyi's approach is composed of two central ideas. First, to get at the uncertainty involving the players' payoff functions, we imagine that each player could be one of several different \textit{types}. A type is possible version of a player possessing specific attributes relevant to their how they select strategies, what their payoff functions are and what they believe about the other players. Types represent possibilities concerning \textit{other players} and and as such should be viewed as an expression of one player's ignorance about key attributes of the other players. The details of this ignorance are quantified by means of probability distributions concerning the players' types themselves. 

Here we collect together the key concepts relating to types both \textit{in general} and for the \textit{specific} case of trading, the latter to be used in the sequel.

\begin{enumerate}
    \item A given player is indicated by an integer $i=1, \dots, n$ and let $K_i$ denote the possible types that player $i$ may obtain.

    \item We write $t_i^k$ to denote player $i$'s $k$-th type and the complete set of types for player $k$ is $t_i^1, \dots, t_i^{K_i}$.

    \item The specific number of types for each player and specific values of the parameters are particular to each type of each player may differ from player to player.
\end{enumerate}

\smallsec{General assumptions}

Naturally this raises several important questions concerning where types come from, what players know about other players' types and how they come to know it. As a starting point we record the following.

\begin{enumerate}
    \item Each player knows what its type is, but is ignorant of what every other players' type is.

    \item The set of types for a given player is \textit{common knowledge} among the other players. That is, while players are specifically ignorant of which \textit{specific} type the other players are, they are aware of the set of \textit{available} type.

    \item There is a \textit{prior distribution} on the joint set of types. on the set of types that is known to all players and agreed upon by all players. That is, for any combination of types $t_1^{k_1}, \dots, t_n^{k_n}$ there is an associated \textit{joint} probability $P(t_1^{k_1}, \dots, t_n^{k_n})$ of those types simultaneously being the active types and the sum over all possible combinations of possible types is unity. 
\end{enumerate}

\smallsec{Notation and terminology}

We begin with some definitions that we will use throughout the sequel.

\medskip

\begin{defn}[Type profiles \cite{harsanyi1994game}]
\label{def:type-profiles}
Consider an $n$-player game where player $i$ ($i=1,\dots, n$) has one of $K_i$ types. Then we make the following definitions:

\begin{enumerate}
    \item \textit{Type profile:} A \textit{type profile} is an $n$-tuple $y=\types$, associating to each player $i$ ($i=1,\dots, n$) a type $k_i$.  Write $t_i^k\in y$ if player $k$ is type $i$ in the profile $y$.

    \item \textit{Profile set:} For a particular game, we write $\T$ for the set of all type profiles for the game. 

    \item \textit{Restricted profile set:} Fix a player $i$ assume their type is $k$. Write $\T_i^k$ for the set of all type profiles $y_i^k\in \T$ such that $t_i^k\in y_i^k$.
\end{enumerate}
\end{defn}

\medskip

\begin{assumption}[Common prior assumption]
\label{ass:common-prior}
For an $n$-player game with incomplete information, where the types are denoted by $t_i^k$ for $i = 1, \dots, n$ and $k = 1, \dots, K_i$, we assume:

\begin{enumerate}
    \item There exists a probability distribution $P$ over the set of type profiles $y \in \T$, describing the probability $P(y)$ of each type profile $y \in \T$.

    \item The distribution $P$ is \textit{common knowledge} among all players.
\end{enumerate}
\end{assumption}

This is referred to as the \textit{common prior assumption}, a central concept in games with incomplete information\footnote{The common prior assumption has been a subject of controversy and is arguably \textit{least} plausible in the types of situations described in this paper. For detailed discussions and justifications of the assumption, see, for example, \cite{morris1995common} and the references therein, as well as \cite{aumann1987correlated}. We do not explore these issues in depth here.}. 

Under this assumption, each player knows their own type but is ignorant of the types of the other players. However, all players have knowledge of the \textit{possible} types of every other player, as well as the joint, unconditional probability distribution over all types. A useful perspective is to view each player as possessing a \textit{private} piece of information—namely, their own type. A consequence of this private information is that each player may form a \textit{subjective} probability distribution over the types of the other players, conditional on their own type. In other words, given that player $i$ is of type $k$, they can form \textit{conditional} probabilities over the active type profile $ y_i^k $.

\begin{equation}
    \pi_i^k(y_i^k) = P(y_i^k | t_i^k)
\end{equation}

which in simple terms means the probability of type profile $y_i^k\in \T_i^k$ conditional on player $i$ being type $k$.

\smallsec{Two-player games}

We note for later that in the case of two players with $K$ and $M$ types respectively, the common prior assumption means there is a $K\times M$ probability matrix of the form:

\begin{equation}
    \label{eq:common-prior-2}
    P   = \pmat{
        p_{11} & p_{12} & \cdots & p_{1M} \\
        p_{21} & p_{22} & \cdots & p_{2M} \\
               &        & \ddots &        \\
        p_{K1} & p_{22} & \cdots & p_{KM}                
    }
\end{equation}

and we write the conditional that when player 1 is type $k$ that player 2 is type $m$ as:

\begin{equation}
    \label{eq:cond-prob-1}
    \pi_1^k(m) := Pr(t_2^m | t_1^k) = p_{km} / \sum_{m'=1}^M p_{km'}
\end{equation}

where $p_{km}$ is the probability player 1 and 2 \textit{simultaneously} being types $k$ and $m$. By the same token we have

\begin{equation}
    \label{eq:cond-prob-2}
    \pi_2^m(k) := Pr(t_1^k | t_2^m) = p_{km} / \sum_{k'=1}^K p_{k'm}
\end{equation}

Note that from player 1's perspective it cannot know which type it will show up as, but it does know that it will show up as one of types $m=1, \dots, M$. Now let $s_2^m\in \S_2^m$ denote whatever strategy player 2 trades when it shows up as Type $m$ (this is unknown to player 2, of course). 

\smallsec{Transition to Complete Information Games with Imperfect Information}

We now describe how an $n$-player game with one or more possible types for each player can be translated into a game with complete but \textit{imperfect} information\footnote{A game with \textit{imperfect information} refers to one in which there are multiple rounds of play, and players lack full knowledge of the moves made in previous rounds. Although this definition is not directly utilized in this paper, it is included here for completeness.}. The key idea is that the game begins with an initial step in which "chance" determines the active type of each player. Following this, players proceed according to the rules of the game, their active types, and their conditional probability distributions regarding the types of the other players.

The role of chance varies depending on the type of incomplete information game being played. Broadly, these games can be divided into two categories. The first category includes games where player types are determined by a public event—an event originating from a trusted source in which all relevant information is equally known to all players. For example, in poker, each player is dealt a hand of cards before the betting begins. In this setting, a player's type is defined by their hand (the specific cards they are dealt), their available funds for betting, and the admissible actions they can take based on their "stack" and the applicable rules.

The second category includes games where player types are determined by non-public events, for example, strategic interactions like trading in competition, where a player's active type is shaped by economic, political, and psychological forces. In these cases, the determination of types involves a mixture of public and private factors that add complexity to the analysis.

In the first category, the common prior assumption is easy, even trivial, to accept. For example, in poker, a player's type is their hand, and assuming a fair and well-shuffled deck, the prior distribution over possible hands is objectively defined. However, in trading, where players are firms, a firm's type may include its market impact model, target quantities, and potential restrictions on admissible trading strategies. In such scenarios, agreeing on a common prior is more challenging due to the private and subjective nature of the underlying factors.

However, even in games like poker, where a player's type initially appears straightforward (e.g., the hand they are dealt), the situation becomes more complex if the player's type is expanded to include attributes such as playing tendencies, levels of aggression, and propensity to bluff. This demonstrates that even games with seemingly objective type definitions can involve subtleties that challenge the assumption of a common prior.

In any case, the introduction of a "move by chance" that determines player types at the start of the game effectively transitions the game into one of complete information. This transformation ensures the existence of Nash equilibria, providing a robust and satisfactory solution concept for analyzing games that might otherwise seem intractable.

\medskip

\subsection{Strategy profiles and payoff functions}
\label{sec:gen-payoff-fns}

We begin by defining the \textit{correct} analogy to a strategy in a game with incomplete information. In a game with complete information a player faces a known adversary, which in this context means that a given player knows the \textit{type} of their adversary. However, in a game with an incomplete information a player faces one of many possible types, each showing up with a given probability. Further, in a game with complete information, a player can determine the optimal action to take conditional on their adversary chooses to play a specific strategy. In a game with incomplete information, a player must choose the optimal strategy given their adversary \textit{would} play a particular strategy when showing up as a particular type. We leverage this idea to define the notion of a \textit{strategy profile}.

\medskip

\begin{defn}[Strategy profiles \cite{harsanyi1967games}]
\label{def:strategy-profiles-gen}
Consider an $n$-player game with incomplete information and a type profiles $\T$. A strategy profile is defined in the following stages:

\begin{enumerate}
    \item \textit{Strategy profile for a player:} When player $i$ ($i=1, \dots, n$) is type $k$ then write $s_i^k$. Write 
    
    \begin{equation}
     s_i^* = (s_i^{1}, \dots, s_i^{K_i})   
    \end{equation}
    
    for the tuple of $K_i$ strategies thus represented. We call this a \textit{strategy profile} for player $i$ and this represents a collection of \textit{conditional} strategies, one for each type $k=1, \dots, K_i$ that player $i$ \textit{might} show up as\footnote{This was called a \textit{normalized} strategy in \cite{harsanyi1967games}.}.

    \item \textit{Strategy profile for all types in the game:} Given a strategy profile $s_i^*$ for each player $i$ ($i=1, \dots, n$), define $s^* = (s_1^*, \dots, s_n^*)$ as the combined strategy profile for all players across all types.

    \item \textit{Strategy profile restricted to a type profile:} For a strategy profile $s^*$ and a specific type profile $y \in \T$, we denote by $s^*(y)$ the strategies corresponding to the $n$ types in $y$.
\end{enumerate}

The third point warrants additional explanation. A strategy profile $s^*$ represents a comprehensive mapping of all strategies to be employed by all players, contingent on their respective types. When we write $s^*(y)$, we are isolating the subset of strategies to be played, conditional on the types of each player being those specified in the type profile $y$.
\end{defn}

\medskip

\begin{defn}[Conditional payoff functions \cite{harsanyi1994game}] 
\label{def:conditional-payout-fns}
Assume that $y\in \T$ is the set of active types for our game. When player $i$ is type $k$ we define the payoff to player $i$ when they are type $k$ as given by:

\begin{equation}
    \label{eq:conditional-payoff-fns-abs}
        V_i^k(s^*(y), y), \qquad \text{if $t_i^k\in y$}
\end{equation}

The meaning of this is that when player $i$ is of type $k$ and this is their \textit{active type} (because $t_i^k\in y$) there is a payoff associated with the other players playing the strategies from there profiles associated with the type profile $y\in \T$. The payoff is conditional for two reasons:

\begin{enumerate}
    \item The payoff occurs to player $i$'s type $k$ \textit{only if} they are the active type; and

    \item The payoff is conditional on the strategy profile $s^*$ and its restriction to the type profile $y$.
\end{enumerate}
\end{defn}

\medskip

Now fix a player $i$ and a type profile $y \in \T_i^k$ (that is, a type profile that fixes player $i$'s type at $k$). Then we write $s_y^*$ for the tuple of strategies $s_j^{k_j}$, $j\ne i$ and such that $s_j^{k-1}\in s_j^*$.

\medskip

\begin{defn}[Expected payoff function]
\label{def:expected-payoff-fn}
Given a strategy profile $s^*$ the expected payoff to player $i$ when they are type $k$ is defined as 

\begin{equation}
    E_i^k(s^*) := \sum_{y\in \T_i^k} \pi_i^k(y) V_i^k(s^*(y), y)
\end{equation}
\end{defn}

\medskip

With these definitions in place we are in a position to define a game with Bayesian players and their equilibrium. To start let $G$ be an $n$-player game with incomplete information. Define types $t_i^k, k=1,\dots, K_i$ for $i=1,\dots, n$ and let $\T$ be the set of all type profiles for players in $G$. Write $\S_i^k$ for the set of admissible strategies available to player $i$ when represented by type $k$.

\medskip

\begin{defn}[Game with "Bayesian" players]
A game with Bayesian players is a game with $n$ players derived from $G$ is a new game $G^*$ defined as follows. First, there is a distribution $P$ over $\T$ that is common knowledge to all players. For this game the strategies and payoffs are as follows:

\begin{enumerate}
    \item \textit{Strategies:} Each player is equipped with strategies $s_i^*$ consisting of a tuple $(s_i^1, \dots, s_i^{K_i}$) such that $s_i^k$ is an admissible strategy for player $i$ when represented as type $k$, We assemble all of the strategy profiles into a single strategy profile $s^*$. Thus each player's strategy in this game is a strategy profile $s_i^*$ consisting of admissible strategies in the original game; 

    \item \textit{Chance play:} Prior to any play by the players, "chance" selects an active type for each player; and

    \item \textit{Payoffs:} Once the active types for each player are selected,  write $k$ for the active type of player $i$ and $t_i^k$ for the type. The payoff to player $i$ with respect for the strategy profile $s^*$ is $E_i^k(s^*)$, the expected payoff with respect to all possible type profiles that contain $t_i^k$.
\end{enumerate}
\end{defn}

\medskip

With this definition in place we are able to define a Nash equilibrium in the new "Bayesian game" derived from the original game.

\medskip

\begin{defn}[Bayes equilibrium]
\label{def:bayes-equi-general}
Let $G^*$ be as above and $s^*$ a strategy profile consisting of profiles $s_1^*, \dots, s_n^*$ for players $i=1, \dots, n$. Then $s^*$ is a Bayes equilibrium if the following holds for all players $i$ with active type $k$. If we modify $s^*$ to $s_\mi^*$ for any player $i$ by replacing $s_i^*$ with a different strategy profile $s_i^{*,1}$ and

\begin{equation}
    E_i^k(s_\mi^*) \le E_i^k(s^*)
\end{equation}
\end{defn}

Given the notational complexity of the setup, we lay the entire setup out for the case of two firms next.

\subsection{Bayesian equilibrium for two firms}

We now specialize type theory and Bayesian equilibrium to trading in competition and restrict ourselves to the case of two firm, illustrating how the theory works in the simpler case. Let $i=1, 2$ denote the firms and let $K$ and $M$ be the number of admissible types for firms 1 and 2 respectively. For a given firm $i$, $\mi$ will denote the "other firm", that is, if $i=1$, $\mi=2$ and for $i=2$, $\mi=1$. We write $\pi_i^k(m)$ for the probability that firm $\mi$ is type $m$ conditional on firm $i$ being type $k$. Given this, the strategy profiles for firms as in definition \ref{def:strategy-profiles-gen} now simplify to a $K$-tuple of admissible strategies for firm 1 and an $M$-tuple for firm 2 as follows:

\begin{equation}
    \begin{split}
        s_1^* = (s_1^1, \dots, s_1^K) \\
        s_2^* = (s_2^1, \dots, s_2^M)
    \end{split}
\end{equation}

and a strategy profile is simply the concatenation of $s^*$ of $s_1^*$ and $s_2^*$:

\begin{equation}
    s^* = (s_1^1, \dots, s_1^K, s_2^1, \dots, s_2^M)
\end{equation}

Type profiles as in Definition \ref{def:type-profiles} now simplify an assignment of a specific type to each firm. In simple terms, a type profile can be thought of as a specification of the \textit{actual type} of each firm that "shows up" to trade. There are $K \times M$ type profiles, and these can be written in lexicographical order as follows:

\begin{equation}
    y_1 = (t_1^1, t_2^1),\, y_2 = (t_1^1, t_2^2),\, \dots,\, y_M = (t_1^1, t_2^M),\, y_{M+1} = (t_1^2, t_2^1), \dots, y_{K\cdot M}  = (t_1^K, t_2^M)
\end{equation}

And we write $\T$ for the set of all profiles, $\T = \set{y_1, y_2, \dots, y_{KM}}$. Meanwhile, suppose that firm 1's active type is 1, then the set of type profiles \textit{restricted to} type $t_1^1$ is as follows:

\begin{equation}
    \T_1^1 = \set{ (t_1^1, t_2^1),\, (t_1^1, t_2^2),\, (t_1^1, t_2^3), \dots,\, (t_1^1, t_2^M) }
\end{equation}

Next, when we have a strategy profile for all types restricted to a specific type profile, this isolates one type for each strategy. So that a strategy profile for $s^* = (s_1^1, \dots, s_1^K, s_2^1, \dots, s_2^M)$ restricted to $y=(t_1^k, t_2^m)$ is $(s_1^k, s_2^m)$. Given this we will write conditional payout functions (see Definition \ref{def:conditional-payout-fns}) in terms of the specific types involved, namely $V_1^k(s_1^k, s_2^m)$ and $V_2^m(s_2^m, s_1^k)$. Given these we define the expected payoff functions (see Definition \ref{def:expected-payoff-fn}) as:

\begin{equation}
    \label{eq:expected-payoff-1}
    E_1^k(s_1^k, s_2^*) := \sum_{m=1}^M \pi_1^k(m) V_1^k(s_1^k, s_2^m)
\end{equation}

where $s_2^* = (s_2^1, \dots, s_2^M)$. For player 2 the conditional payout function is given as:

\begin{equation}
    \label{eq:expected-payoff-2}
    E_2^m(s_2^m, s_1^*) := \sum_{k=1}^K \pi_2^m(k) V_2^m(s_1^k, s_2^m)
\end{equation}

with $s_2^* :=(s_2^1, \dots, s_2^M)$; $s_1^*$ and $s_2^*$ are called \textit{strategy profiles}. Given all of this we may define a Bayesian equilibrium in for two players in a manner equivalent to Definition \ref{def:bayes-equi-general}.

\medskip

\begin{defn}[Specific best-response to a strategy profile] 
\label{def:best-response-concrete}
When studying trading as a game of complete information, the notion of a best-response strategy is clear and intuitive. If, for example, firm $2$ trades the strategy $s_2$ then player 1's \textit{best-response} strategy to $s_2$ is the strategy that minimizes the expected cost when trading in competition with $s_2$\footnote{This is discussed in detail in \cite{chriss2024optposnbldg}}. In a game with incomplete information firm $i$ ($i=1, 2$) faces one of many possible adversaries depending on which of the possible active types for firm $\sm i$ shows up. 

In this case, to define a best response strategy firm 1 (say), we have to contend with the uncertainty as to which firm 2 active type shows up. Firm 1 faces not a strategy but a \textit{strategy profile} $s_2^* = (s_2^1,\dots, s_2^M)$ (see Definition \ref{def:strategy-profiles-gen}). Recalling that $S_i^k$ is the set of admissible strategies for firm $i$, type $k$. We say the specific strategy $s_1^k$ is the best-response to the strategy profile $s_2^*$ if its expected value $E_1^k(s_1^k, s_2^*)$, \cref{eq:expected-payoff-1}, is maximal among all strategies $s\in S_1^k$: 

\begin{equation}
\label{eq:best-resonse-concrete}
    E_1^k(s_1^k, s_2^*) = \max_{s \in S_1^k} E_1^k(s, s_{2}^*)
\end{equation}

Likewise $s_2^k$ is a best-response to $s_1^*$ if

\begin{equation}
\label{eq:best-resonse-concrete-2}
    E_2^k(s_2^k, s_2^*) = \max_{s \in S_2^k} E_2^k(s, s_{1}^*)
\end{equation}
\end{defn}

\medskip

\begin{defn}[Best-response strategy profile] 
\label{def:best-response-profile}
Now consider firm 1 from firm 2's perspective. Suppose that firm 2 believes that firm 1 is trading the strategy profile $s_1^*$ and wants to produce a strategy profile $s_2^*$ each component of which is a best-response to $s_1^*$. Such a strategy profile is called a \textit{best-response strategy profile}. A similar definition applies for the best response profile for firm 1 versus a known profile $s_2^*$ for firm 2. What is the meaning of this?

In that firm 2 (say) knows what type it is, why does it need a complete best-response profile? The answer has to do with firm 1's perspective. Since firm 1 does not know what firm 2's active type will be, it is necessary for firm 2 to have a best-response strategy for each of its potential types. 
\end{defn}

\medskip

\begin{defn}[Bayesian equilibrium]
\label{def:bayes-equi}
Strategy profiles $s_1^*, s_2^*$ for firms 1 and 2 respectively are a \textit{Bayes equilibrium set of strategies} if $s_1^*$ is a best-response profile to $s_2^*$ and $s_2^*$ is \textit{simultaneously} a best-response to $s_1^*$. In other words:

\begin{subequations}
    \begin{align}
        s_1^k &= \max_{s_1\in \S_1^k} E_1^k(s_1, s_2^*), \qquad k=1, \dots, K \label{eq:firm1-eqi}\\
        s_2^m &= \max_{s_2\in \S_2^m} E_2^m(s_2, s_1^*), \qquad m=1, \dots, M \label{eq:firm2-eqi}
    \end{align}
\end{subequations}
\end{defn}

and a Bayesian equilibrium strategy is given by 

\begin{equation}
    \label{eq:bayes-equi}
    \begin{split}
    s_1^* &= (s_1^1, \dots, s_1^K) \\    
    s_2^* &= (s_2^1, \dots, s_2^M) \\    
    \end{split}
\end{equation}

In the next section we derive Bayesian equilibria for strategies in the two-firm case.

\section{Trading in competition with "Bayesian" firms}
\label{sec:trading-in-comp-bayesian}

We now introduce games with "Bayesian" players in the context of trading in competition, utilizing the Bayesian games framework described in Section \ref{sec:games-incomplete}. We begin with the two-firm case, as it is conceptually identical to the general case apart from notational differences. The more complex notation required for an arbitrary number of firms is introduced in Section \ref{sec:more-than-two-firms}. Examples are deferred to Section \ref{sec:two-firm-examples-1}.

This section is largely self-contained and demonstrates how to apply the concepts introduced in Section \ref{sec:games-incomplete} to analyze and solve games arising in the context of trading in competition. Along the way, we outline the relevant details and requisite assumptions. Section \ref{sec:market-impact-models} reviews the market impact models used in the analysis, which build on the Almgren-Chriss framework (see \cite{almgren2001optimal}) and extend the models for trading in competition with complete information (see \cite{chriss2024equilibria}). Section \ref{sec:key-assumptions} examines the key assumptions underlying these models, discussing their strengths and limitations.

The purpose of this section is to get to the trading strategies in \ref{sec:solving-bayesian}, which details the process for solving Bayesian equilibrium using Harsanyi's theory of types. Once the foundational mathematical framework is established, the key results are readily produced using a straightforward application of the Euler-Lagrange equation, requiring no special or new mathematics. That being said, a new result is presented in Section \ref{sec:non-strategic-firm}, where we show how to derive strategies for two firms trading in competition when there is a possibility of a third, \textit{non-strategic} firm entering the market. This extends the preceding sections to account for scenarios involving non-strategic participants.

\subsection{Trading and Game Theory}
\label{sec:trading-and-game-theory}

Portfolio managers often trade in anticipation of a future catalyst expected to move a stock's price in their favor. Their goal is to maximize profit while minimizing the market impact of their own trades and those of competing firms. The combined trading activity of all firms influences both current and future prices, directly affecting the profitability of their trades.

To minimize the adverse effects of their trades, firms typically build their positions over time, spreading out trades before the catalyst. The timing and execution of these trades constitute the firm's \textit{trading strategy}. The game-theoretic problem is to determine how firms should choose their strategies while accounting for the actions of competing firms. This problem was analyzed in \cite{chriss2024equilibria} and \cite{chriss2024optposnbldg}, where trading was modeled as a game with complete information.

We now introduce the standard terminology that will be used throughout the rest of this paper:

\begin{enumerate}
    \item \textit{Trading games} refer to games that arise when two or more market participants compete for the same stock during the same time period.

    \item A \textit{firm} is one of the players in a trading game.

    \item A \textit{market impact model} is a mathematical framework that describes how trades in a stock affect its current and future price (see Section \ref{sec:market-impact-models}).

    \item A \textit{trading strategy} specifies the timing and size of trades over a given period, designed to acquire a predetermined quantity of stock within a set timeframe. A trading strategy defines precisely \textit{how} the stock is to be acquired.

    \item The \textit{implementation cost} of a trading strategy is the difference between the stock's price at the start of the strategy (referred to as the \textit{arrival price}) and the average price per share paid while executing the strategy.
\end{enumerate}

With these definitions in pace we describe the key \textit{implicit} assumptions used to derive Nash equilibrium solutions in \cite{chriss2024equilibria} and \cite{chriss2024optposnbldg}.

\begin{enumerate}
    \item Every firm uses the same market impact model in order to select their strategies selection. Specifically, market impact model used are the "Almgren-Chriss" model in \cite{almgren2001optimal}. The precise details are described in the sequel, see Section \ref{sec:market-impact-models} and \cref{eq:cost-eqns}.
    
    \item All firms not only use the same functional form for the market impact model but they also use the same \textit{market impact parameter}, as is described in Section \ref{sec:market-impact-models}.

    \item Every firm has a specific \textit{target quantity } they wish to acquire during the course of trading.

    \item Each firm seeks to minimize their \textit{implementation cost} of acquiring their target quantity and the implementation cost is measured by the market impact model.
\end{enumerate}

Finally, it is assumed that 1, 2 and 3 above are \textit{common knowledge} among the firms trading. That is, each firm knows that every other firm knows that each firm is using the Almgren-Chriss model with a specific market impact parameter, every firm knows every other firm's target quantity and every firm knows that every other firm knows their target quantity, and so on.  

In the sequel we will analyze how to incorporate uncertainty concerning the above parameters into trading in competition. To do so, we need to be somewhat more precise and mathematical.

\subsection{Mathematical Formulation}
\label{sec:mathematical-formulation}

We now formalize the definitions provided above. A trading strategy $s(t)$ in a stock is a function of time, $t$, defined over the interval $[0, 1]$. The strategy begins at time $t=0$ and ends at time $t=1$, specifying the holdings of the stock $s(t)$ at each time $t$. Key assumptions include:

\begin{enumerate}
    \item Time is scaled so that all trading strategies occur over the interval $[0, 1]$, regardless of the actual duration in physical time.

    \item Unless otherwise stated, $s(0)=0$, meaning the strategy starts with no stock holdings and ends with a total quantity $s(1)$.

    \item Trading strategies are assumed to be "pre-specified," meaning they are fully determined before trading begins. As a result, the strategies analyzed in this paper are \textit{ex ante} trading "plans," with the associated limitations discussed in the introduction.
    
    \item The quantity $s(1)$ is referred to as the \textit{target quantity} of the strategy.

    \item If $s(t)$ is differentiable at a time $t \in [0, 1]$, we denote its derivative by $\dot{s}(t)$. This represents the \textit{instantaneous rate of trading} at time $t$.
\end{enumerate}

Let $X_t$ represent a model for the evolution of the stock price, incorporating the impact of the strategy $s$. This model generally takes the form:

\begin{equation}
    X_T = X_0 + \int_0^T f(\dot s) \dt + \text{random noise} 
\end{equation}

Now assume there are $n$ firms trading in competition with strategies $s_1, \dots, s_n$ and that each firm believes that the correct market impact model is $f_i$. Write $s=\sum_i s_i$ and call this the \textit{aggregate trading} of firms $i=1, \dots, n$. 

\medskip

\begin{defn}[Implementation cost]
\label{def:imp-cost}
Given $n$ firms in competition with strategies $s_1,\dots, s_n$ and $s$ as above. Then the implementation cost to firm $i$ with respect to the market impact function $f_i$ is 

\begin{equation}
    \label{eq:imp-cost}
    \text{Cost}(s_i; s, f_i) = \int_0^1 f(\dot s) \du 
\end{equation}
\end{defn}

We note that the implementation cost to player $i$ depends on both the strategies employed by every other firm (and summarized in the strategy $s$) \textit{and} on the market impact function $f_i$ of firm $i$. It is a central premise of this paper that firms trading in competition seek to minimize their implementation costs in light of their view of the correct market impact function $f_i$ and what strategies the other firms will select.

\subsection{Market impact models}
\label{sec:market-impact-models}

With these in place, we proceed to define the conditional payoff functions discussed in general in Section \ref{sec:gen-payoff-fns}, \cref{eq:conditional-payoff-fns-abs} for the specific case of trading in competition. For this we need a market impact model that describes how trading in a stock impacts current and future prices. For this paper we use the "Almgren-Chriss" model developed in \cite{almgren2001optimal} which was also the basis for \cite{chriss2024equilibria} in describing trading games with complete information\footnote{Though we use these specific models here, in principle any models may be used, subject to overcoming whatever computational and analytic difficulties that may arise.}. 

Briefly, this model is as follows. For players 1 and 2 trading strategies $s_1$ and $s_2$ respectively (we omit the type notation for the moment) the cost of trading for players 1 and 2 are given by

\begin{equation}
    \label{eq:cost-eqns}
    \begin{split}
        C_1 &= \int_0^1 \big(\dot s_1(t) + \dot s_2(t)\big) \dot s_1(t) + \kappa \big(s_1(t) + s_2(t)\big) \dot s_1(t) \dt \\
        C_2 &= \int_0^1 \big(\dot s_1(t) + \dot s_2(t)\big) \dot s_2(t) + \kappa \big(s_1(t) + s_2(t)\big) \dot s_2(t) \dt        
    \end{split}
\end{equation}

Here $C_1$ is the implementation cost to player 1 assuming a market impact parameter of $\kappa$, and $C_2$ is the cost to player 2. In these expressions, the time derivatives $\dot s_1(t)$ and $\dot s_2(t)$ should be regarded as the quantity of trading over an infinitesimal amount of time so that if the price per share of the stock is $x$ then the amount paid in that infinitesimal period of time is $x \, \dot s(t)$. The total cost consists of the sum of two components:

\begin{enumerate}
    \item \textit{Temporary} impact due to the aggregate rate of trading by both firms (given by $\dot s_1 + \dot s_2$) times the rate of trading of the firm whose cost we are measuring ($\dot s_1$ for firm 1 and $\dot s_2$ for firm 2). 

    \item \textit{Permanent} impact due to the aggregate \textit{cumulative} trading both firms (given by $s_1(t) + s_2(2)$). This relates to the \textit{informativeness} of the trading, see \cite{chriss2024equilibria}. 
\end{enumerate}

We will use the form of the cost functions to define conditional payoff functions in a moment, but first we make some key points regarding their form and some of their potential implications.

\smallsec{Notation}

We utilize the following notation throughout our discussion. Throughout we restrict discussions to two firms, called simply firm 1 and firm 2. 

\begin{enumerate}
    \item \textit{Admissible strategies:} Write $\S_1^k$ for the set of admissible strategies available to firm 1 when its active type is $k$. Similarly write $\S_2^m$ for admissible strategies for firm 2 when its active type is $m$;
    
    \item {Specific admissible strategies:} Write $s_1^k \in \S_1^k$ and $s_2^m\in \S_2^m$ for specific \textit{admissible strategies} for firms 1 and 2 when their active types are $k, m$ respectively;

    \item \textit{Market impact coefficient:} Write $\kappa_1^k, \kappa_2^m$ for the market impact parameters used by firms 1 and 2 (respectively) when their active types are $k, m$ respectively; and

    \item \textit{Target quantities:} Write $f_1^k, f_2^m$ for the target quantities for firms 1 and 2 when their active types are $k, m$ respectively. 

    \item \textit{Strategy profiles:} As in Definition \ref{def:strategy-profiles-gen} a strategy profile for firm 1 is a strategy specification for each of its possible active types. This is written as an $M$-tuple of strategies $s_1^* = (s_1^1, s_1^2, \dots, s_1^M)$. Similarly, a strategy profile for firm 2 is a a $K$-tuple of strategies $s_2^* = (s_2^1, \dots, s_2^K)$,
\end{enumerate}

\smallsec{Discussion}

When translating mathematical models into practice, subtle details can sometimes lead to unexpected results. Below, we outline several key points for future reference:

\begin{enumerate}
    \item \textit{Simultaneous trading is additive:} The temporary cost components in \cref{eq:cost-eqns} as stated imply that when firms trade \textit{at the same time}, only their aggregate trading affects instantaneous market impact. For example, if firm 1 trades $x$ shares per unit of time and firm 2 trades $-x$ shares over a time interval $[t, t+\Delta t]$, their trades fully offset each other, resulting in zero market impact. This property is a feature of the model, referred to as \textit{simultaneous trading is additive}.

    \item \textit{Permanent impact is additive:} The permanent impact cost components depends only on the \textit{aggregate} cumulative trading of the firms involved. This means that if one firm is buying while another is selling, the net effect on future prices depends solely on the net amount traded by both firms combined.

    \item \textit{Indistinguishability:} The previous two points imply that the "market" does not distinguish between two firms trading such that their combined net trading is $x$ shares. For instance, if firm 1 trades $a$ shares and firm 2 trades $b$ shares such that $a+b=x$, the market impact is identical to a scenario where a single firm trades $x$ shares.

    \item \textit{The market impact model is common knowledge:} This assumption is central to the analysis. The conditional payoff functions for the "Bayesian trading game" are defined based on the specific market impact model and cost functions in \cref{eq:cost-eqns}. This explicitly assumes that all participants use this model and that it is common knowledge among them.

    \item \textit{Participants seek to minimize costs:} Building on the previous point, we assume that (i) all firms aim to minimize their costs based on this model and (ii) the only source of disagreement among firms is the market impact parameter. However, there are many possible market impact models, and uncertainty about which model each firm uses could itself be a significant factor.
\end{enumerate}

As a simple example, imagine that there are two "acceptable" market impact models and it is known that approximately 80\% of firms use model 1 and 20\% use model 2. Then one could establish as a baseline two "umbrella" types, one for model 1 and one for model  2 with specific types under each umbrella to fill out additional uncertainty concerning the parameters used to implement those models. 

\subsection{Parameter uncertainty \textit{is} payoff function uncertainty}
\label{sec:uncertainty-payoff}

Trading in competition has the following main categories of uncertainty when some number of firms are in competition to buy or sell the same stock over the same period of time. 

\begin{enumerate}
    \item There may be uncertainty about the precise mathematical form of the market microstructure model (e.g., market impact model) employed by each firm in estimating the implementation cost of strategies (see Definition \ref{def:imp-cost}). 

    \item There may be uncertainty about the precise quantity each firm seeks to trade, i.e., their target quantities (see Section \ref{sec:mathematical-formulation}). 

    \item There may be uncertainty about the payoff functions of each firm. That is, if firms $i=1, \dots, n$ select strategies $s_1, \dots, s_n$ the value to firm $j$ of these strategies, written $V_j(s_1, \dots, s_n)$ may be uncertain. 

    \item There may be uncertainty about what the feasible strategies available to each firm are, that is, what the \textit{strategy spaces} of each firm are. For example, some firms may be constrained to never shorting stocks they wish to acquire, while others may not be allowed to \textit{overbuy} or may be limited to only a certain amount of overbuying.
\end{enumerate}

\subsection{Key assumptions}
\label{sec:key-assumptions}

We now describe they key concepts of Bayesian games specialized to trading games in the form of a set of assumptions. 

\medskip

\begin{assumption}[Admissible strategies]
We begin by assuming that every game has specified set of pure strategies available to each player. In trading games, the strategies are trading strategies specified by functions of time\footnote{The details of such strategies are given in \cite{chriss2024optposnbldg} and \cite{chriss2024equilibria}}. The set of \textit{valid} strategies are the \textit{admissible trading strategies}. We assume that the sets of admissible strategies for firms 1 and 2 respectively consist of the set of twice-differentiable functions 

\begin{equation}
    s: [0, 1]\to \R 
\end{equation}

such that $s(0)=0$, where each $t\in [0, 1]$ represents time with $t=0$ being the "start time" (of trading) and $t=1$ the end time. This assumption amounts to $\S_1^k, \S_2^m$ are the set of twice differentiable functions from $[0, 1]$ to the reals that vanish at zero, for all $k, m$.
\end{assumption}

Next we port the common prior assumption here.

\medskip

\begin{assumption}[Common prior assumption, see Assumption \ref{eq:common-prior-2}]
There is a probability matrix $P$ whose $(k, m)$ entry represents the joint probability of $t_1^k$ and $t_2^m$. This means that not only are the possible types for each firm \textit{common knowledge} but the unconditional probabilities are as well. 
\end{assumption}

\medskip

\begin{assumption}[Conditional payoff functions, see Definition \ref{def:conditional-payout-fns}]
\label{ass:conditional-payoff}

Assume that firm 1's active type is $k$ and firm 2's is $m$ and that firm 1 trades the strategy $s_1^k$ when its active type is $k$, and firm 2 trades $s_2^m$ when its active type is $m$. Then when firm 1 competes with the strategy $s_2$ and firm 2 competes with the strategy $s_1$. we write $V_1^k(s_1^k, s_2$ and $V_2^m(s_1, s_2^m)$ for the \textit{conditional payoff functions} for firms 1 and 2 respectively. These are given by the negative values of the cost functions in \cref{eq:cost-eqns}:

\begin{subequations}
    \begin{align}
        V_1^k(s_1^k, s_2) &= -\int_0^1 (\dot s_1^k + \dot s_2)\dot s_1^k + \kappa_1^k (s_1^k + s_2) \dot s_1^k \dt \label{eq:cond-payoff-firm1-2firms} \\
        V_2^m(s_1, s_2^m) &= -\int_0^1 (\dot s_1 + \dot s_2^m)\dot s_2^m + \kappa_2^m (s_1 + s_2^m) \dot s_2^m \dt \label{eq:cond-payoff-firm2-2firms}
    \end{align}
\end{subequations}
\end{assumption}

\medskip

The functions $V_1^k, V_2^m$ are \textit{payoff} functions in the following sense. These are the values that trades would like to strategically maximize because in doing so this minimizes costs which in turns maximizes the profit involved in their trade. In the sequel we will revert to using cost functions, negatives of equations \eqref{eq:cond-payoff-firm1-2firms}-\eqref{eq:cond-payoff-firm2-2firms}.

\smallsec{Key features of conditional payoff functions} 

\begin{enumerate}
    \item The conditional payoff functions are conditional in the sense that they express what firms 1 and 2 pay \textit{conditional} what their active types are. For example, $V_1^k(s_1^k, s_2)$ is the payoff that firm 1 will receive when its active type is 1 and it trades strategy $s_1^k$ and firm 2 trades strategy $s_2$.

        \item Implicit in the conditional payoff functions is that they are \textit{subjective} in that the market impact parameter for a given firm $i$ and active type $k$, the market $\kappa_i^k$ is based on the firm's \textit{belief} that this is the correct market impact parameter\footnote{The subjectivity goes deeper than this because assumption \ref{ass:conditional-payoff} says, in effect, that all firms use the same mathematical formulation for market impact in general. In fact, it is possible that each firm uses a different market impact function describing how trading impacts market prices.}.

    \item The parameter $\kappa_i^k$ is referred to as the \textit{market impact coefficient}. The form of the conditional payoff functions means that each firm's active type has a belief concerning what $\kappa$ is. These types are used by each active type to select their trading strategies. However, once they trade, both firms are subject to the \textit{same} market conditions. There is only one market impact parameter governing trading costs. We call this the \textit{real market impact parameter}. This will be discussed more in Section \ref{sec:performance-evaluation} when we discuss trading cost evaluation.
\end{enumerate}

In the sequel we will also switch from payoff maximization to cost minimization. We call this an assumption to flag the fact that as noted in Section \ref{sec:gen-payoff-fns} the types themselves should contemplate uncertainty concerning payoff functions themselves. In effect, then, we are making the assumption that it is \textit{common knowledge} that each firm uses the same basic market impact function and that all of the uncertainty in the trading game lies in what the parameters and target quantities are.     

\subsection{Bayesian equilibrium}
\label{sec:solving-bayesian}

The expected payoff functions as discussed in Section \ref{sec:bayesian-equilibrium} are now straightforward to write down using the conditional payoff functions \eqref{eq:cond-payoff-firm1-2firms}-\eqref{eq:cond-payoff-firm2-2firms}:

\begin{subequations}
\begin{align}
    E_1(s_1^k, s_2^*; \kappa_1^k) &= -\sum_{m=1}^M \pi_1^k(m) \int_0^1 (\dot s_1^k + \dot s_2^m) \dot s_1^k + \kappa_1^k(s_1^k + s_2^m) \dot s_1^k \dt \label{eq:exp-cost1}\\
    E_2(s_2^m, s_1^*; \kappa_2^m) &= -\sum_{k=1}^K \pi_2^m(k) \int_0^1 (\dot s_2^m + \dot s_1^k) \dot s_2^m + \kappa_2^m(s_2^m + s_1^k) \dot s_2^m \dt \label{eq:exp-cost2}
\end{align}    
\end{subequations}

Where $\pi_i^k(m)$ is the \textit{conditional probability distributions} as in \eqref{eq:cond-prob-1}-\eqref{eq:cond-prob-2}. These expressions are the negatives of the expected costs each firm incurs implementing their specific strategy against an unknown (conditional) strategy of their adversary. To solve for Bayesian equilibrium, according to equations \eqref{eq:firm1-eqi}-\eqref{eq:firm2-eqi} we must \textit{simultaneously} maximize equations \eqref{eq:exp-cost1}-\eqref{eq:exp-cost2}. Negating both of the latter expressions for $E_1, E_2$ we want to find a strategies $s_1^k, s_2^m$ that minimize \cref{eq:exp-cost1} and \cref{eq:exp-cost2} at the same time. We  use the Euler-Lagrange equation in the standard way to conclude that for all $k=1, \cdots, K$ and $m=1, \cdots, M$ we \textit{must} have:

\begin{subequations}
    \begin{align}
        \ddot s_1^k &= -\frac{1}{2} \sum_{m=1}^M \pi_1^k(m)(\ddot s_2^m + \kappa_1^k \dot s_2^m) \\
        \ddot s_2^m &= -\frac{1}{2} \sum_{k=1}^K \pi_1^m(k)(\ddot s_2^k + \kappa_2^m \dot s_2^k) 
    \end{align}
\end{subequations}

These equations can be re-written as a system of first-order equations in matrix form that is both more generalizable and easier to solve. Start by introducing the variables $v_i^j = \dot s_i^j$ so that $\dot v_i^j = \ddot s_i^j$.  This immediately implies:

\begin{subequations}
\begin{align}
        \dot v_1^k &= -\frac{1}{2} \sum_{m=1}^M \pi_1^k(m)(\dot v_2^m + \kappa_1^k v_2^m) \\
        \dot v_2^m &= -\frac{1}{2} \sum_{k=1}^K \pi_2^m(k)(\dot v_2^k + \kappa_2^m v_2^k)     
\end{align}
\end{subequations}

After moving all terms involved with $v$ to the left-hand side and all terms involving $\dot v$ to the right-hand side the equations become:

\begin{subequations}
\begin{align}
        \dot v_1^k + \frac{1}{2} \sum_{m=1}^M \pi_1^k(m) \dot v_2^m &= -\frac{1}{2} \sum_{m=1}^M \pi_1^k(m) \kappa_1^k v_2^m \label{eq:firm1-eq}\\
        \dot v_2^m + \frac{1}{2} \sum_{k=1}^K \pi_2^m(k)\dot v_2^k  &= -\frac{1}{2} \sum_{k=1}^K \pi_2^m(k) \kappa_2^m v_2^k \label{eq:firm2-eq}
\end{align}
\end{subequations}

Our aim is to combine equations \eqref{eq:firm1-eq}-\eqref{eq:firm2-eq} into a single matrix equations of the form 

\begin{equation}
    \label{eq:vdot-matrix-eqn}
    A \dot v = -B v
\end{equation}

where $v:= (v_1 \, v_2)^\top$ with $v_1$ and $v_2$ defined in \cref{eq:v1-v2-definitions} and $A$ and $B$ being the matrices of coefficients arising from \cref{eq:firm1-eq} and \cref{eq:firm2-eq}. 

\begin{equation}
    \label{eq:v1-v2-definitions}
    v_1 = \pmat{v_1^1 \\ \vdots \\ v_1^K} \quad v_2 = \pmat{v_2^1 \\ \vdots \\ v_2^M}
\end{equation}

From the coefficients of the left-hand-side of \cref{eq:firm1-eq} and \cref{eq:firm2-eq} respectively we define the matrices $A_1$ and $A_2$ as follows:
 
\begin{equation}
    \label{eq:mat-A1}
    \begin{split}
        A_1 = \pmat{
            1 & 0 & \cdots & 0 & \frac{1}{2} \pi_1^1(1) & \cdots & \frac{1}{2}\pi_1^1(M) \\ 
            0 & 1 & \cdots & 0 & \frac{1}{2} \pi_1^2(1) & \cdots & \frac{1}{2}\pi_1^2(M) \\ 
              &   & \vdots &   &                        & \vdots &            \\
            0 & 0 &        & 1 & \frac{1}{2} \pi_1^K(1) & \cdots & \frac{1}{2}\pi_1^K(M)
        }, \qquad K \times (K + M)
    \end{split}
\end{equation}

Similarly $A_2$ is the $M \times (K+M)$ matrix representing the coefficients of the left-hand side of \cref{eq:firm2-eq}:

\begin{equation}
    \label{eq:mat-A2}
    \begin{split}
        A_2 = \pmat{
            \frac{1}{2} \pi_2^1(1) & \cdots & \frac{1}{2}\pi_2^1(K) & 1 & 0 & \cdots & 0 \\ 
            \frac{1}{2} \pi_2^2(1) & \cdots & \frac{1}{2}\pi_2^2(K) & 0 & 1 & \cdots & 0 \\ 
              &   & \vdots &   &                        & \vdots &            \\
            \frac{1}{2} \pi_2^M(1) & \cdots & \frac{1}{2}\pi_2^M(K) & 0 & 0 & \cdots & 1
        }, \qquad M \times (K + M) 
    \end{split}
\end{equation}

and finally define:

\begin{equation}
    \label{eq:A-mat-def}
    A := \pmat{A_1 \\ A_2}, \qquad    
\end{equation}

We next look at the right-hand sides of \cref{eq:firm1-eq} and \cref{eq:firm2-eq} to define $B_1$ and $B_2$ as follows:

\begin{equation}
    \label{eq:mat-B1}
    \begin{split}
        B_1 = \pmat{
            0 & 0 & \cdots & 0 & \frac{1}{2} \pi_1^1(1) \kappa_1^1 & \cdots & \frac{1}{2}\pi_1^1(M)\kappa_1^1 \\ 
            0 & 0 & \cdots & 0 & \frac{1}{2} \pi_1^2(1)\kappa_1^2 & \cdots & \frac{1}{2}\pi_1^2(M)\kappa_1^1 \\ 
              &   & \vdots &   &                        & \vdots &            \\
            0 & 0 &        & 0 & \frac{1}{2} \pi_1^K(1) \kappa_1^K & \cdots & \frac{1}{2}\pi_1^K(M)\kappa_1^K
        }, 
    \end{split}
\end{equation}

and

\begin{equation}
    \label{eq:mat-B2}
    \begin{split}
        B_2 = \pmat{
            \frac{1}{2} \pi_2^1(1)\kappa_2^1 & \cdots & \frac{1}{2}\pi_2^1(K)\kappa_2^1 & 0 & 0 & \cdots & 0 \\ 
            \frac{1}{2} \pi_2^2(1)\kappa_2^2 & \cdots & \frac{1}{2}\pi_2^2(K)\kappa_2^2 & 0 & 0 & \cdots & 0 \\ 
              &   & \vdots &   &                        & \vdots &            \\
            \frac{1}{2} \pi_2^M(1)\kappa_2^M & \cdots & \frac{1}{2}\pi_2^M(K)\kappa_2^M & 0 & 0 & \cdots & 0
        }
    \end{split}
\end{equation}

and define

\begin{equation}
    \label{eq:B-mat-defn}
    B := \pmat{B_1 \\ B_2}  \quad (K+M) \!\times\! (K+M)
\end{equation}

Using this notation we may re-write \eqref{eq:firm1-eq} and \eqref{eq:firm2-eq} as

\begin{equation}
    A \dot{v} = -B v.
\end{equation}

which becomes:

\begin{equation}
    \label{eq:two-firm-equi}
    \dot{v} = -A^{-1} B v
\end{equation}

where $v(t) = [v_1(t), v_2(t)]^\top$. If we write $M:= -A^{-1}B$ then the entire system of equations for $\dot v$ may be written as:

\begin{equation}
    \label{eq:two--firm-equi-with-M}
    \dot{v} = M v
\end{equation}

\smallsec{Robustness: using the normal form equations}

We quickly note that in some circumstances $A$ may be non-singular or nearly singular. In this instance we may express the equations in \textit{normal form} as:

\begin{equation}
    A^T A \dot v = -A^T B v
\end{equation}

and therefore we may write $\dot v$ in terms of the normal-form equation:

\begin{equation}
    \label{eq:two-firm-equi-robust}
    \dot v = -(A^T A)^{-1} A^T B v 
\end{equation}

Therefore in the case where $A$ is singular (or \textit{could be} singular) we can use the \textit{alternative} definition of $M$ given my

\begin{equation}
    M:= -(A^T A)^{-1} A^T B
\end{equation}

In the case $A$ is non-singular this is equivalent to \ref{eq:two-firm-equi} because as is well-known $(A^T A)^{-1} A^T = A^{-1}$ (when $A$ is non-singular )\footnote{This is because when $A$ is non-singular then $(A^T A)^{-1}= A^{-1} (A^T)^{-1}$ then $(A^T A)^{-1} A^T = A^{-1} (A^T)^{-1} A^T = A^{-1}$.}. In the next section we use this to explicitly solve for the relevant strategies. In general we will proceed without reference to the normal-form equations but note that these should be employed when $A$ may not singular or nearly singular. 

\smallsec{Computation of solutions}

To begin, by \cref{eq:first-order-soln} $v(t)$ may be expressed as follows letting $v_0 := v(0)$:

\begin{equation}
    \label{eq:s_dot_expr}
    v(t) = e^{M t} v_0
\end{equation}

We set our aim to find the most straightforward description of the matrix $C=-A^{-1} B$ as possible. To begin we construct matrices $\Pi_1$, $\Pi_2$ from the sub-blocks of equations \eqref{eq:mat-A1} and \eqref{eq:mat-A2} consisting of conditional probabilities:

\begin{equation}
\label{eq:pi-mats}
    \Pi_1 = \frac{1}{2}\pmat{\pi_1^1(1) & \cdots & \pi_1^1(M) \\ 
                             \pi_1^2(1) & \cdots & \pi_1^2(M) \\ 
                                        & \vdots &            \\
                             \pi_1^K(1) & \cdots & \pi_1^K(M)}, \quad
    \Pi_2 = \frac{1}{2}\pmat{\pi_2^1(1) & \cdots & \pi_2^1(K) \\ 
                             \pi_2^2(1) & \cdots & \pi_2^2(K) \\ 
                                        & \vdots &            \\
                             \pi_2^M(1) & \cdots & \pi_2^M(K)}
\end{equation}

Clearly, $\Pi_1$'s dimensions are $K \times M$ and $\Pi_2$'s $M \times K$. Then the matrix $A$ in \cref{eq:A-mat-def} becomes:

\begin{equation}
    \label{eq:A-compact}
    A = \pmat{I & \Pi_1 \\ \Pi_2 & I}
\end{equation}

We may represent $B$ compactly as well after taking into account the presence of the $\kappa_i^k$. To deal with this define diagonal matrices $D_1$ and $D_2$ respectively as:

\begin{equation}
\label{eq:d1-d2}
    D_1 := \pmat{\kappa_1^1 \\ & \kappa_1^2 \\ & & \ddots \\ & & &  \kappa_1^K}, \quad
    D_2 := \pmat{\kappa_2^1 \\ & \kappa_1^2 \\ & & \ddots \\ & & &  \kappa_1^M}
\end{equation}

Then the matrix $B$ becomes:

\begin{equation}
    \label{eq:B-compact}
    B = \pmat{0 & D_1 \Pi_1 \\ D_2 \Pi_2 & 0}
\end{equation}

Our aim now is to use the expressions for $A$ and $B$ in \eqref{eq:A-compact} and \eqref{eq:B-compact} to construct a compact expression for $e^{Ct}$, $C = -A^{-1} B$. Note that by construction the matrix $I - \Pi_1 \Pi_2$ does not have $1$ as an eigenvalue and therefore is invertible. Given this we may write:

\begin{equation}
    \label{eq:A-inv}
    A^{-1} = \pmat{(I - \Pi_1 \Pi_2)^{-1} & -\Pi_1 (I - \Pi_2 \Pi_1)^{-1} \\
            -\Pi_2 (I - \Pi_1 \Pi_2)^{-1} & (I - \Pi_2 \Pi_1)^{-1}}
\end{equation}

where $0$ is the $K \times K$ (top row) and $M \times M$ matrix of zeros. From this it is possible to write $C = -A^{-1} B$ compactly as follows:

\begin{align}
C = -A^{-1} B &= \pmat{
            -(I - \Pi_1 \Pi_2)^{-1} & \Pi_1 (I - \Pi_2 \Pi_1)^{-1} \\
            \Pi_2 (I - \Pi_1 \Pi_2)^{-1} & -(I - \Pi_2 \Pi_1)^{-1}}
    \pmat{0 & D_1 \Pi_1 \\D_2 \Pi_2 & 0}  \nonumber \\[1em]
    &= \pmat{
        \Pi_1 (I - \Pi_2 \Pi_1)^{-1} D_2 \Pi_2 & -(I - \Pi_1 \Pi_2)^{-1} D_1 \Pi_1 \\
        -(I - \Pi_2 \Pi_1)^{-1} D_2 \Pi_2 & \Pi_2 (I - \Pi_1 \Pi_2)^{-1} D_1 \Pi_1
        }  \label{eq:C-expr-1}
\end{align}

Thus \cref{eq:C-expr-1} is a compact representation of $C$. Now our aim is to find a set of Bayesian optimal strategies $s_1, s_2$, as defined above. Let $s$ be the $K + M$ dimensional strategy vector consisting of the $K$ possible strategies for firm 1 and $M$ for firm 2:

\begin{equation}
    \label{eq:strat-vector}
    s = (s_1^1, \cdots, s_1^K, s_2^1, \cdots, s_2^M)^\top, \qquad s_i^k: [0, 1]\to \R
\end{equation}

Now, assuming both firms start with no stock and at time $t=1$ firm 1's target quantities are given by the $K$ values $f_1^k$, $k=1,\cdots, K$ and firm 2's are given by the $M$ values $f_2^m$, $m=1, \cdots, M$, so that

\begin{equation}
    \label{eq:boundary-1}
    f = (f_1^1, \cdots, f_1^K, f_2^1, \cdots, f_2^M)^\top
\end{equation}

Using \cref{eq:v_0} we provide an expression for $v_0 := v(0)$:

\begin{equation}
    \label{eq:v0-defn-1}
    v_0 = M (e^{M} - I)^{-1} f
\end{equation}

and setting $M = -A^{-1} B$ the general solution to such a linear differential equation yields an expression for $s(t)$:

\begin{equation}
    \label{eq:soln-st}
    s(t) = (e^{Mt} - I) v_0
\end{equation}

Note that \cref{eq:soln-st} provides a complete description of a two-firm Bayesian equilibrium of the two \textit{specific strategies}: $s_1^* = (s_1^1, \dots, s_1^K)$ and $s_2^*=(s_2^1,\dots, s_2^M)$ satisfying the boundary conditions in \cref{eq:boundary-1}. We also note that from \cref{eq:s_dot_expr} we immediately have:

\begin{equation}
    \label{eq:soln-sdot}
    \dot s(t) = e^{Mt} v_0
\end{equation}

\subsection{Two firms competing in the presence a non-strategic firm}
\label{sec:non-strategic-firm}

To illustrate the tremendous generality and flexibility of types, we consider the following situation:

\begin{enumerate}
    \item There are two firms in competition with one another.

    \item There is the \textit{possibility} of a third, \textit{non-strategic} firm trading the same stock as firms 1 and 2 over the same period of time. The non-strategic firm is still in competition with the other two firms but employs none of the game-theoretic logic of strategy selection.

    \item We call the first two firms the \textit{strategic} firms and the add to their type data a third item which describes their belief concerning the non-strategic firms target size (see Section \ref{sec:types}).

    \item Thus the type data for each firm in this situation is a triple consisting of the each types belief in what the market impact coefficient is, what their target size is and what the non-strategic firm's target size is.
\end{enumerate}

Below we will analyze how to apply the machinery of Bayesian games to this situation. First, however, we show that the above setup is in some sense a natural generalization of the Almgren-Chriss model. 

\medskip

\smallsec{Optimal execution in the presence of a non-strategic firm}

The results in this section can be applied to optimal execution settings as in \cite{almgren2001optimal} when a single firm is executing a large trade in the presence of a second firm that is \textit{non-strategic}. In this setup, assuming no risk the loss function would be

\begin{equation}
    L(t) = \int_0^1 (\dot s(t) + b) \dot s(t) + \kappa(s(t) + b\cdot  t)\dot s(t) \dt 
\end{equation}

where $b\cdot t$ is the non-strategic firm's strategy (a risk-neutral strategy with target size $b$). Using the Euler-Lagrange equation we find the optimal strategy is given by the equations:

\begin{equation}
    2 \ddot s(t) + \kappa b = 0
\end{equation}

Assuming that the strategic firm's target quantity is $f$ then the its optimal strategy is:

\begin{equation}
    s(t) = -\frac{1}{2}\kappa b t^2 + a_1 t + a_2 
\end{equation}

then we see that $a_1=f+\frac{1}{2}\kappa b$ and $a_2=0$. In other words, the optimal strategy is an \textit{eager} strategy whose degree of curvature is proportional to the product of the market impact parameter used and the target size of the non-strategic investor. The rest of this section may be regarded as a generalization of this result to multiple firms. We note that this is precisely the best-response to a risk-neutral firm as described in \cite{chriss2024optposnbldg}.
    
\smallsec{Two strategic firms in competition with a non-strategic firm}

We refer to the first two firms as the \textit{strategic} firms, indicating that they will select their strategies \textit{game-theoretically}. The third firm that trades without taking into account the other firms will be referred to as the \textit{non-strategic} firm. 

\begin{enumerate}
    \item When referring to firm $i=1, 2$ we will write $\smi$ for "the other firm", that is $\smi=2$ when $i=1$ and $\smi=1$ when $i=2$.

    \item We will write $K$ and $M$ for the number of types for firm 1 and 2 respectively. 
    
    \item Each possible active type of firms 1 and 2 believes that the non-strategic firm will trade a risk-neutral strategy with a specific target size. We write $b_i^k$ for firm $i$, type $k$'s belief about the target size of the non-strategic firm. Note that it is possible that some of the $b_i^k$ are equal to zero meaning that firm $i$, type $k$ believes there not a non-strategic investor.

    \item The type data for firm $i$ when its active type is $k$ is a triple $(\kappa_i^k, f_i^k, b_i^k)$, where $\kappa_i^k$ is firm $i$'s market impact coefficient, $f_i^k$ is its target size and $b_i^k$ represents the target size that firm $i$ (when its active type is $k$) believes the non-strategic firm will trade.

    \item There is a common prior probability matrix $P$, see \cref{eq:common-prior-2}, such that $p_{km}$ is the joint probability that firm 1's active type is $k$ and firm 2's is $m$. Note that unlike before, each firm's type \textit{includes} their view of the non-strategic firm's target size. 

    \item From $P$ we have $\pi_i^k(m)$ is the \textit{conditional probability} that if firm $i$ is type $k$ then firm $\smi$ (the other firm) is type $m$.
\end{enumerate}

We note that it is possible and straightforward to extend the analysis we do so that some firms types believe that the non-strategic firm trades some other strategy other than the risk-neutral strategy and our analysis directly extends to this case so long as the other other strategy's functional form is fixed and known ahead of time. In the case of the risk neutral strategy its functional form is $s(t) = b_i^k t$ for firm $i$, type $k$. Therefore it is clear that the expected payoff function, see \cref{eq:expected-payoff-1}, for firms $1$ and $2$ when their active types are $k$ and $m$ respectively are given by: 

\begin{align}
    \text{Firm 1} &= -\sum_{m=1}^{M} \pi_1^k(m) \int_0^1 (\dot s_1^k + \dot s_2^m + b_1^k) \dot s_1^k + \kappa_i^k (s_1^k + s_2^m + b_1^k \, t) \dot s_1^k \dt \\
    \text{Firm 2} &=-\sum_{k=1}^{K} \pi_2^m(k) \int_0^1 (\dot s_1^k + \dot s_2^m + b_2^k) \dot s_2^m + \kappa_i^k (s_1^k + s_2^m + b_1^k \, t) \dot s_2^m \dt
\end{align}

Now using the Euler-Lagrange equation we see that the strategy that minimizes the expected payoff for firm $1$ (firm 2 is analogous) when its active type is $k$, is given by the following differential equation in terms of the $s_\smi^k$:

\begin{equation}
    \ddot s_1^k + \frac{1}{2} \sum_{m=1}^{M} \ddot s_2^m = -\frac{1}{2} \sum_{m=1}^{M} \pi_1^k(m) \kappa_1^k \dot s_2^m - \frac{1}{2}\kappa_i^k b_1^k
\end{equation}

Now write $A$ as in \cref{eq:A-compact}, B as in \cref{eq:B-compact}. Also $f$ for the target sizes as in \cref{eq:boundary-1} and $s$ for the combined strategy vector as in \cref{eq:strat-vector}. Now write $b$ for the combined vector of $b_i^k$, the non-strategic sizes. 

\begin{equation}
    \label{eq:non-strategic-sz-vec}
    b := (b_1^1, b_1^2, \dots, b_1^K, b_2^1, \dots b_2^M)^\top
\end{equation}

and 

\begin{equation}
    \label{eq:kappa-vec-def}
    \kappa := (\kappa_1^1, \dots, \kappa_1^K, \kappa_2^1, \dots, \kappa_2^M)^\top   
\end{equation}

Now write $\star$ for \textit{element wise} product and $v = \dot s$ as in Section \ref{sec:solving-bayesian}. it is straightforward to see that analogous to \cref{eq:vdot-matrix-eqn} we have:

\begin{equation}
    A \dot v = -B v - \frac{1}{2} \kappa \star b
\end{equation}

We may now solve this equation \textit{robustly} using the normal-form equation using the pseudo-inverse $(A^T A)^\dagger$ for the pseudo-inverse of $A^T A$, then 

\begin{equation}
    \label{eq:normal-form-v-goliath}
    \dot v = -(A^TA)^\dagger A^T B v - \frac{1}{2}  (A^TA)^\dagger A^T\kappa \star b
\end{equation}

To compute a concrete solution for $s(t) = \int v$ we first define $M = -(A^TA)^\dagger A^T B$ and 

\begin{equation}
    \label{eq:c-vec-defn}
    c := -\frac{1}{2}(A^TA)^\dagger A^T \kappa \star b
\end{equation}

so that \cref{eq:normal-form-v-goliath} becomes:

\begin{equation}
    \label{eq:normal-form-v-goliath-2}
    \dot v = -M v + c
\end{equation}

Now write $v_0 := v(0)$ we have:

\begin{equation}
    v(t) = e^{M t} \left( v_0 + M^{-1} c \right) - M^{-1} c
\end{equation}

Which means we have:

\begin{equation}
    \label{eq:v0-goliath}
    v_0 = \big(M^{-1} (e^M - I )\big)^{-1} \big(f + M^{-1} c - M^{-1} (e^M - I ) M^{-1} c\big)
\end{equation}

and we have an expression for the entire strategy vector $s(t) = \int_0^t v(\tau) \,\text{d$\tau$}$ in terms of $v_0$ and $c$.

\begin{equation}
    \label{eq:strategy-goliath}
    s(t) = M^{-1} \left(e^{M t} - I \right) \left(v_0 + M^{-1} c \right) - t M^{-1} c
\end{equation}

We summarize the meaning of \cref{eq:strategy-goliath} as follows writing $K:=K_1$ and $M:=K_2$.

\begin{enumerate}
    \item The strategy $s(t)$ consists of one component for each firm and type. Thus it is a vector of the form $s=(s_1^1, s_1^2, \dots, s_1^K, s_2^1,\dots, s_2^M)^\top$ where $s_i^k$ is the trading strategy form firm $i$ when its active type is $k$. Thus the Bayesian game strategy (see Definition \ref{def:bayes-equi}) for firms 1 and 2 are given by:
    \begin{equation}
        \begin{split}
            s_1^* &= (s_1^1, \dots, s_1^K) \\
            s_2^* &= (s_2^1, \dots, s_2^M)
        \end{split}
    \end{equation}

    \item The target size of firm $i$, type $k$ is $f_i^k$ and the target size vector is $f=(f_1^1, \dots, f_1^K, f_2^1, \dots, f_2^M)^\top$. 

    \item The vector $c$ is as in \cref{eq:c-vec-defn} and encapsulates the intensity of trading of the non-strategic firm in terms of the products of the market-impact parameters and the \textit{believed} target sizes $\kappa_i^k, b_i^k$.

    \item The matrix $A$ is defined as in \cref{eq:A-mat-def} and \cref{eq:A-compact}. The matrix $M$ is as defined in \cref{eq:two--firm-equi-with-M} when $A$ is non-singular and as in \cref{eq:two-firm-equi-robust} when $A$ may be singular.
    
    \item Firm $i$, type $k$ believes that there is a non-strategic investor trading the strategy $s(t) = b_i^k t$ so that $s(1)=b_i^k$, the target size that this firm/type believes the non-strategic investor is trading toward.

    \item The vector $c$ is the element-wise product $c = -\frac{1}{2} \kappa * b$ where $b$ is the target size vector such that $b_i^k$ is the target size that firm $i$ and type $k$ believes the non-strategic firm will trade and $\kappa * b$ is the element-wise product. 

    \item Each firm and type trades taking into account its belief that there is a non-strategic firm that is trading a risk-neutral strategy of some size, and also takes into account that each of the other firm's possible types believes that there is a non-strategic trader trading a risk-neutral strategy of some size.     

    \item One can see that when $b=0$ (that is $b_i^k=0$ for all $i, k$) then $\kappa * b = 0$ and therefore $c=0$. Examining \cref{eq:v0-goliath} this shows $v_0 = (e^M - I )^{-1} M f = M(e^M - I)^{-1}f$. 
\end{enumerate}

\section{Strategy performance evaluation}
\label{sec:performance-evaluation}

Once we compute optimal strategies, we would like to be able to \textit{evaluate} their performance by computing the cost functions . To begin with, we will compute cost of trading each strategy on their own terms, namely, we will compute the cost of a given firm's strategy assuming they "show up" as a given type and (in the case of two firms) the other firm shows up as a particular type. For example, in the case of two firms and three types per firm, if firm one shows up as type one then there are three \textit{possible types} that firm two may show up as, and we can compute the realized cost of each as well as the \textit{subjective probability} from firm one's point of view, of the cost of trading.  

\subsection{Implementation cost with two firms in competition}
\label{sec:imp-cost-2-comp}

Consider the standard setup with two firms, 1 and 2, having $K$ and $M$ types, respectively. The strategy vector is defined as $s(t) = (s_1^1(t), \cdots, s_1^K(t), s_2^1(t), \cdots, s_2^M(t))^\top$, as in \cref{eq:strat-vector}, subject to the boundary conditions $s(0) = 0$ and $s(1) = f$, where $f$ is a vector of target quantities for each firm and each type.

Suppose the active types for firms 1 and 2 are $k$ and $m$, respectively. We aim to compute the predicted trading costs that will be incurred by each firm, as measured by their payoff functions, assuming a \textit{common} alpha-decay parameter $\kappa$ meant to represent the "true" market impact . The costs for firms 1 and 2 are then given as follows:

\begin{subequations}
    \begin{align}
        C_1(s_1^k, s_2^m; \kappa) &= \int_0^1 (\dot s_1^k + \dot s_2^m) \dot s_1^k + \kappa  (s_1^k +  s_2^m) \dot s_1^k \dt \label{eq:actual-cost-1} \\
        C_2(s_2^m, s_1^k; \kappa) &= \int_0^1 (\dot s_2^m + \dot s_1^k) \dot s_2^m + \kappa  (s_2^m +  s_1^k) \dot s_2^m \dt \label{eq:actual-cost-2}
    \end{align}
\end{subequations}

We call the market impact parameter $\kappa$ used in computing $C_1$ and $C_2$, the \textit{evaluation market impact parameter} or simply the \textit{evaluation} $\kappa$. The underlying idea of this is that once firm 1 shows up as Type 1 and firm 2 shows up as Type 2 everything is concrete and there is a specific alpha-decay parameter at work. We would like to compute the costs to each firm for trading their \textit{actual} types. To compute the above integrals it is enough to compute $\dot s$ which by definition is the same as $v$. 
Then to compute $s(t)$  and $\dot s(t)$ we use \cref{eq:soln-st} and \cref{eq:soln-sdot}:

\begin{subequations}
    \begin{align}
        s(t)  &= (e^{Mt} - I)M^{-1} v_0 \label{eq:s_expr}\\
    \dot s(t) &= e^{M t} v_0  \label{eq:s_dot_expr} 
    \end{align}
\end{subequations}

and were given the boundary conditions $s(0)=0, s(1)=f$ we have $v_0 = M(e^M - I)^{-1} f$. In matrix form we write:

\begin{equation}
\renewcommand{\arraystretch}{1.66}    
\pmat{\dot s_1^1(t) \\ \vdots \\ \dot s_1^K(t) \\ \dot s_2^1(t) \\ \vdots \\ \dot s_2^M(t)} = e^{M t} v_0
\end{equation}

\subsection{Cumulative costs over time}
\label{sec:cumulative-costs}

In the sequel we will analyze not just the total actual implementation cost of a strategy but also how that cost evolve as the strategy is traded. To do this, is a simple matter of changing the upper limit of integration from 1 to some time $t < 1$. Thus we write $C_1(t; s_1^k, s_2^m, \kappa$ to for the \textit{cumulative} cost of trading for firm 1 trading strategy $s_1^k$ when its active type is $k$ against firm 2 trading strategy $s_2^m$ when its active type is $m$ and when the actual market impact parameter is $\kappa$. We write $C_2(t; s_2^m, s_1^k$ for the analogous function for firm 2. To calculate these requires nothing more than performing the following integrations.

\begin{subequations}
    \begin{align}
        C_1(t; s_1^k, s_2^m, \kappa) &= \int_0^t (\dot s_1^k + \dot s_2^m) \dot s_1^k + \kappa  (s_1^k +  s_2^m) \dot s_1^k \dt \label{eq:cum-actual-cost-1} \\
        C_2(t; s_2^m, s_1^k, \kappa) &= \int_0^t (\dot s_2^m + \dot s_1^k) \dot s_2^m + \kappa  (s_2^m +  s_1^k) \dot s_2^m \dt \label{eq:cum-actual-cost-2}
    \end{align}
\end{subequations}

\subsection{The case when there are more than two firms}
\label{sec:more-than-two-firms}

The case where there are more than two firms is surprisingly similar to that of two firms. The key thing is to define the $A$ and $B$ matrices in equations \eqref{eq:A-compact} and \eqref{eq:B-compact} respectively. Recall that both $A$ and $B$ in the two firm case consist of two blocks:

\begin{equation}
    \label{eq:A-compact-gen}
    A = \pmat{I & \Pi_1 \\ \Pi_2 & I}, \quad     B = \pmat{0 & D_1 \Pi_1 \\ D_2 \Pi_2 & 0}
\end{equation}

where $D_1$ and $D_2$ are defined in equation \eqref{eq:d1-d2} while $\Pi_1$ and $\Pi_2$ are defined in \eqref{eq:pi-mats}. Thus each block in the rows of $A$ represent a different player and the number of rows in a block is equivalent to the number of types for the player. Now, each row in A (respectively, $B$) represents a specific player $i$ when represented by type $k$. The number of columns in the given block is therefore given by the number of type profiles in the restricted set $\T_i^k$ (see Definition \ref{def:type-profiles}). 

\section{Two-firm equilibrium examples}
\label{sec:two-firm-examples-1}

In this section, we provide detailed examples of two-firm Bayesian equilibria with multiple types, building on the results of Sections \ref{sec:trading-in-comp-bayesian} and \ref{sec:performance-evaluation}. The aim is to illustrate how the introduction of multiple types (and the resulting uncertainty) influences the strategies firms adopt. We focus on the case of two firms, each with two possible types, trading in competition. The analysis examines how variations in key parameters affect the strategies selected by each firm.

\medskip

\begin{table}[h!]
\centering
\renewcommand{\arraystretch}{1.25} 
\begin{tabular}{lll}
\hline
 & \multicolumn{1}{c}{\text{Type 1}} & \multicolumn{1}{c}{\text{Type 2}} \\
\hline
\text{Firm 1}& $t_1^{1}=(1.0, 3.0)$ & $t_1^{2}=(3.0, 5.0)$ \\
\text{Firm 2}& $t_2^{1}=(2.0, 7.0)$ & $t_2^{2}=(15.0, 5.0)$ \\
\hline
\end{tabular}
\caption{Type data for two firms. Each type $t_i^k=(\kappa_i^k, f_i^k)$ consists of a market impact parameter $\kappa_i^k$ and a target size $f_i^k$. For example, firm 1's type 1 has $\kappa_1^1=1.0$ and $f_1^1=3.0$.}
\label{tab:two-firms-2-types-1}
\end{table}

\medskip

\begin{figure}[h!]
    \centering
    \includegraphics[width=0.90\linewidth]{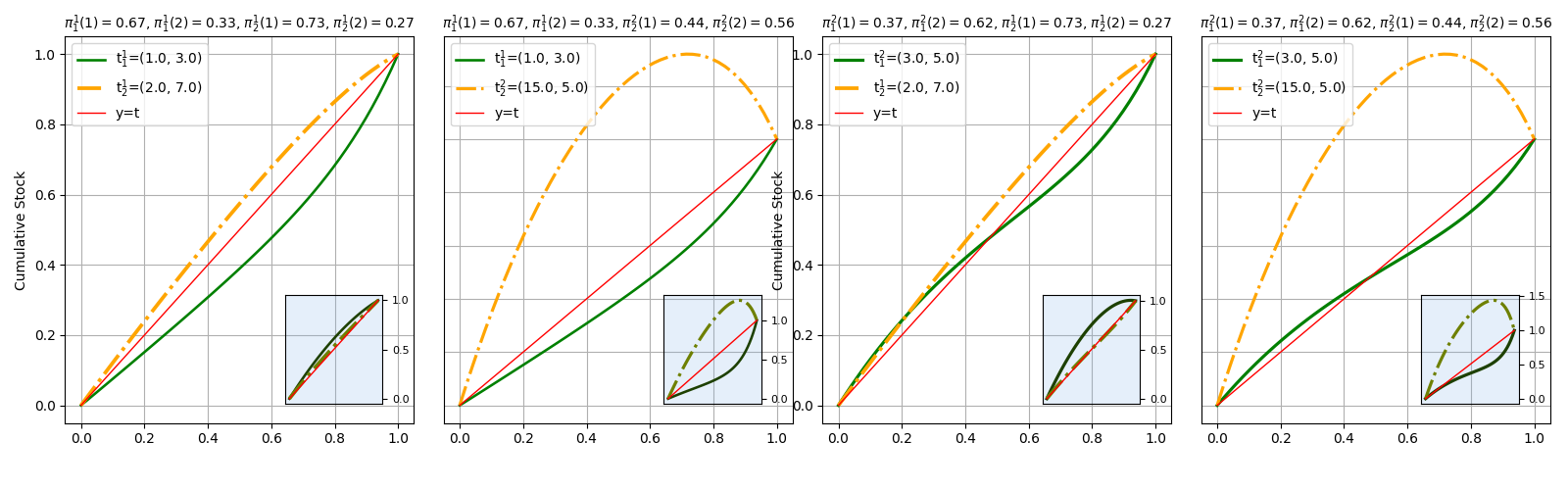}
    \caption{Strategy pairs for two firms with two possible types each using the type data shown in Table \ref{tab:two-firms-2-types-1} and probability matrices in \cref{eq:two-firms-2-types-1}. Each sub-plot shows the trading strategies of firms 1 and 2 when their active types are $k$ and $m$ according to the sub-plot's location. For example, the top-left plot shows the strategies when the active types are both 1. The inset in each plot shows the strategy pair for firms 1 and 2 when there is only one possible type and the type data is the same as the active types in the primary plot. In each plot, solid lines are firm 1's strategy and dash-dotted lines are firm 2's.}
    \label{fig:two-firms-2-types-1}
\end{figure}

Equation \eqref{eq:two-firms-2-types-1} displays the common prior matrix $P$ (see Section \ref{sec:types}) and $P_1$ and $P_2$ the conditional probability matrices \cref{eq:cond-prob-1} and \cref{eq:cond-prob-2} respectively. Figure \ref{fig:two-firms-2-types-1} displays the four strategy pairs according to the four possible pairs of active strategies. In the title of each plot are displayed the conditional probabilities from $P_1$ and $P_2$ to facilitate discussion. 

\begin{equation}
\label{eq:two-firms-2-types-1}
P = \begin{pmatrix}
0.40 & 0.20 \\
0.15 & 0.25
\end{pmatrix} \!,\quad P_1 = \begin{pmatrix}
0.67 & 0.33 \\
0.37 & 0.62
\end{pmatrix} \!,\quad P_2 = \begin{pmatrix}
0.73 & 0.27 \\
0.44 & 0.56
\end{pmatrix}
\end{equation}

Figure \ref{fig:two-firms-2-types-1} displays the trading strategies for firms 1 and 2 in a plot, each sub-plot of which shows one possible pair of active types for the firms, with the $(i,j)$ subplot showing the case where firm 1's active type is $i$ and firm 2's is $j$. To demonstrate the difference between complete and incomplete trading in competition, each plot also displays an inset that shows what the strategies would be if the active types in the primary plot were the active types \textit{with certainty}. In terms of Section \ref{sec:bayesian-equilibrium} these plots show the \textit{strategy profiles} (see Definition \ref{def:strategy-profiles-gen}) where $s_1^* = (s_1^1, s_1^2)$ consists of the firm 1 strategies (solid lines) for $t_1^1$ and $t_1^2$ in the top row of plots and $s_2^*$ consists of the firm 2 strategies (dash-dotted lines) in the first column. 

\smallsec{Discussion}

The strategy pairs displayed in Figure \ref{fig:two-firms-2-types-1} immediately demonstrate that when there are even two types, the strategies traded by the active types need not be anything like the corresponding optimal strategies in the case where the active types are the only types. 

A few features of these plots are helpful to point out. First, it is important to keep in mind that each firm has two possible strategies. Each row of the primary plot grid shows the \textit{same} strategy for firm 1 and each column of the plot grid shows the same strategy for firm 2. All of the strategies are also \textit{normalized} so that their target quantity is one, so that these strategies primarily show the strategies shape. However, the thickness of the lines in the plots are adjusted to show the relative target quantity. 

Consider the following differences according to type pairs, referencing the figure. We will write $(k, m)$ for the tuple representing firm 1 when its active type is $k$ and firm 2 when its active type is simultaneously $m$.

\begin{enumerate}
    \item When the active types of firms 1 and 2 are $(1, 1)$ (the leftmost plot in Figure \ref{fig:two-firms-2-types-1}), firm 2 trades a mildly eager strategy (the dash-dotted line in the primary plot) while firm 1 trades a somewhat risk-averse strategy (the green line). Meanwhile, examining the inset which shows the strategies that \textit{would be} traded if firm 1 were Type 1 and firm 2 were Type 2 \textit{with certainty}.

    \item When the active types are (2, 1) (the third plot in Figure \ref{fig:two-firms-2-types-1}), as with the case of active types (1, 1) firm 2 again trades a modestly eager strategy while firm 1 trades an S-shape strategy that closely hugs the risk-neutral one. Meanwhile in the inset, where firm 1 is Type 2 and firm 2 is Type 1 \textit{with certainty} firm 1 trades an eager strategy. 

    \item On the other hand, when active types are (1, 2) (the second plot in Figure \ref{fig:two-firms-2-types-1}), the strategy shapes for firms 1 and 2 are more similar to the case when there is only one possible type per firm. Even in this case, however, the strategies are not precisely the same as in the complete information case. Both strategies in the latter case are more \textit{extreme} in the sense that their farthest deviation from the risk-neutral strategy is greater then when their are two types per firm.
    
    \item Finally when both types are (2, 2) (the rightmost plot in Figure \ref{fig:two-firms-2-types-1}) provides an interesting example of the effect that uncertainty has on trading strategies. Note that in this case firm 2 believes the market impact coefficient is 15.0 and has a target size of 5.0. So why does it trade so aggressively? This seems substantially different than the complete information case of \cite{chriss2024optposnbldg}. The main reason is because firm 2 knows that firm 1 believes the market impact coefficient is either 1 or 3 and therefore is not concerned with the market impact firm 2 will cause. In the complete information case in \cite{chriss2024optposnbldg} firms share a common believe about the market impact parameter. 
\end{enumerate}

It can be challenging to understand why the shapes of optimal strategies are as they are. To begin, we reason through case where the active types of firm 1 and firm 2 are both one as displayed in the leftmost plot of Figure \ref{fig:two-firms-2-types-1} where firm 1's strategy is risk-averse (convex) when there are two possible types but concave when there is one. 

To begin, consider when firm 1's active type is 1 and firm 2's is 2 \textit{with certainty}. In this case, firm 1 has a target size of 3.0 while firm 2's is 7.0, more than twice as large. Moreover, the firms \textit{disagree} on what the correct value of the market impact parameter (a situation which is different from \cite{chriss2024equilibria} in which all firms agree on the value of $\kappa$). firm 1 believes the correct market impact parameter is 1.0 while firm 2 believes it is 2.0. These beliefs themselves are \textit{common knowledge}, that is firm 1 knows that firm 2 knows it believes $\kappa=1.0$ and vice-versa. Each firm chooses its strategy accordingly, and firm 1 trading a smaller size and knowing firm 2 will trade more cautiously given its beliefs about $\kappa$ can trade a modestly eager strategy versus firm 2's close to risk-neutral strategy as we see in the inset of the top-left plot of Figure \ref{fig:two-firms-2-types-1}.

Now we introduce a second type for firm 1 and firm 2. Now firm 1 is facing one of two possible types and firm 1 assigns a \textit{subjective} probability to each as follows: 

\begin{enumerate}
    \item Firm 2's active type is 1 with a probability of 67\%, believes $\kappa$ is 2.0 and has a target size of 7.0. In this case firm 2 trades a modestly eager strategy because it believes $\kappa$ is relatively small and it has a relatively large target size; or

    \item Firm 2's active type is 2 with a probability of 33\%, believes $\kappa$ is 15.0 and has a target size of 5.0. With the very large $\kappa$ of 15.0 and a target size \textit{at least} as large as both of firm 1's active types' target sizes (firm 1's target sizes are 3.0 and 5.0 for types 1 and 2 respectively), firm 2 trades a very eager strategy in order to get ahead of what it believes will be a significant amount of permanent impact.
\end{enumerate}

Now consider how firm 1 must react to these two strategies. firm 1 believes there is a 67\% probability that firm 2 is type 1 and therefore will trade a modestly eager strategy but with a 33\% probability and a very eager strategy, firm 1 benefits from delaying slowing its trading in order to reduce costs near the end. 

\begin{figure}[h!]
    \centering
    \includegraphics[width=1\linewidth]{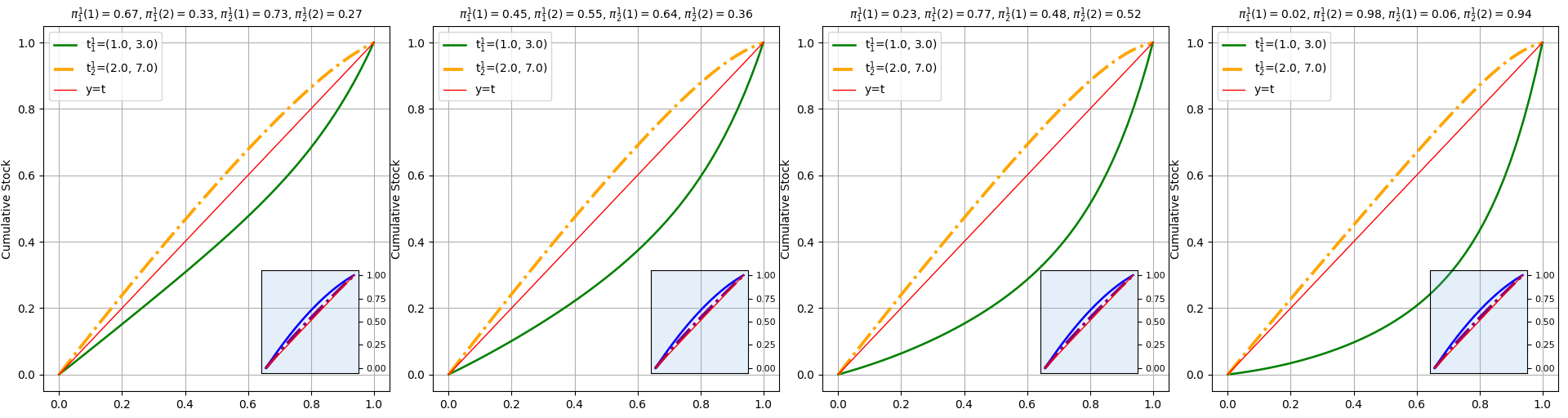}
    \caption{The same strategies as the top-left plot in Figure \ref{fig:two-firms-2-types-1} except that from left to right the firm 1's subjective probability that firm 2 is type 1 is successively decreased.}
    \label{fig:two-firms-2-types-grid-inset-probs}
\end{figure}

In order to demonstrate the impact of firm 1's subjective probabilities on its strategy in Figure \ref{fig:two-firms-2-types-grid-inset-probs} we display firm 1 and firm 2's strategy for precisely the same type data as already displayed in Figure \ref{fig:two-firms-2-types-1} (and as in Table \ref{tab:two-firms-2-types-1}) and with an \textit{almost} identical common prior probability matrix as in \cref{eq:two-firms-2-types-1} except that in each successive plot from left to right we \textit{decrease} the probability in the common prior that both firm 1 and firm 2 have active type 1. 

We display \textit{result} of this in the plot titles by showing the \textit{conditional probabilities}. For example, firm 1's conditional probability of firm 2 being type 1 from left to right are given by $\pi_1^1(1)=67\%$, $\pi_1^1(1)=45\%$, $\pi_1^1(1)=23\%$ and $\pi_1^1(1)=2\%$. Thus in each successive plot firm 1 wen its active type is 1 holds a subjective probability that firm 2 is type 1 that decreases from 67\% to 2\%. Meanwhile firm 2, when its active type is 1, has a subjective belief concerning firm 1's active type being 1 that drops from 73\% to 6\%. The impact on trading is mainly that firm 1 trades increasingly slowly. Let's understand why. Consider the rightmost plot where firm 1's active type is 1 and therefore it believes the market impact coefficient is small at $\kappa=1.0$ and further it believes with close to certainty (94\%) that firm 2's active type is 2 and thus its target size is 5.0 (as opposed to 7.0) with close to certainty. Thus firm 1's beliefs are that $\kappa$ is small and that firm 2 is very likely trading the smaller of its two possible sizes. Moreover, firm 1 believes that firm 2 that the market impact coefficient is very large (because its active type is very likely 2 firm 2 believes $\kappa=15.0$) . With a $\kappa$ that large, firm 1 believes that firm 2 will trade \textit{eagerly} and can benefit from the late selling that firm 2 must do to hit its target size. 

\begin{figure}[h!]
    \centering
    \includegraphics[width=1\linewidth]{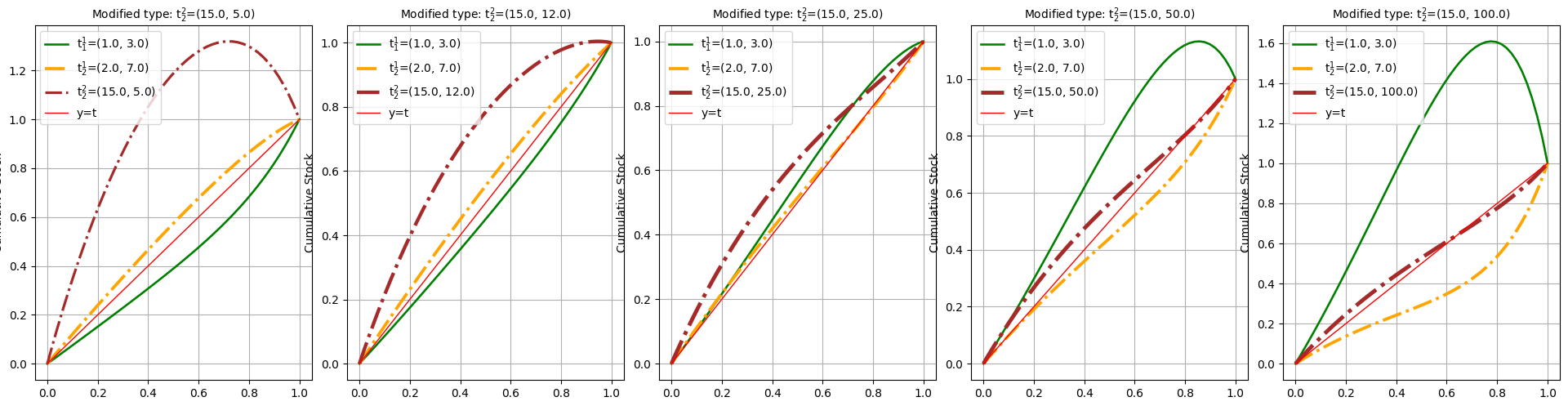}
    \caption{Plots of firm 1 with active type 1 and firm 2 with active type 2 for the same type data as in Table \ref{tab:two-firms-2-types-1} and probabilities as in \cref{eq:two-firms-2-types-1}. In these plots firm 2's type 2 target size is increased as we move from left to right from its baseline of  5.0 to 100.0.}
    \label{fig:two-firms-2-types-grid-inset-sizes}
\end{figure}

Next we demonstrate how changing the target size of a single firm and type impacts the strategies. In Figure \ref{fig:two-firms-2-types-grid-inset-sizes} we start with the same type data and probabilities as in Table \ref{tab:two-firms-2-types-1} and \cref{eq:two-firms-2-types-1} and display the strategies for firm 1 with active type 1 and firm 2 with active type 1 but from left to right increase the \textit{target size} of firm 2's inactive type (type 2) from 5.0 to 100.0. These plots display firm 2's type 2 strategy in the plot as well. In these plots, as before, the subjective probabilities that firm 1 assigns to firm 2's active type being 1 or 2 as 67\% and 33\% respectively. 

One can see from Figure \ref{fig:two-firms-2-types-grid-inset-sizes} that the impact of increasing firm 2's type 2 target size is make its strategy increasingly aggressive until it begins to overbuy at first somewhat and then ultimately significantly, all owing to the possibility that firm 2 \textit{might} have an active type that trades an extremely large target size.

\smallsec{Implementation costs}

Given the strategies above, we discuss their implementation costs as described in Section \ref{sec:performance-evaluation}. We continue to use the type data from Table \ref{tab:two-firms-2-types-1} which we repeat here for convenience.

\begin{table}[h!]
\centering
\renewcommand{\arraystretch}{1.25} 
\begin{tabular}{lll}
\hline
 & \multicolumn{1}{c}{\text{Type 1}} & \multicolumn{1}{c}{\text{Type 2}} \\
\hline
\text{Firm 1}& $t_1^{1}=(1.0, 3.0)$ & $t_1^{2}=(3.0, 5.0)$ \\
\text{Firm 2}& $t_2^{1}=(2.0, 7.0)$ & $t_2^{2}=(15.0, 5.0)$ \\
\hline
\end{tabular}
\end{table}

Each type $t_i^k=(\kappa_i^k, f_i^k)$ consists of a market impact parameter $\kappa_i^k$ and a target size $f_i^k$. For example, firm 1's type 1 has $\kappa_1^1=1.0$ and $f_1^1=3.0$. In Table \ref{tab:two-firms-2-types-1} we display the costs and expected costs as follows.

\begin{enumerate}
    \item Each block of rows displays implementation costs normalized by target size for a different pair of active types. For example, in the first block the active type for firm 1 is 1, and the same goes for firm 2. 
    \item The normalization that in each row firm 1's costs are normalized by firm 1's target size for its active type and the same holds for firm 2. For example, in the First row, the costs associated with firm 1 have been divided by its target size which is 3.0, while firm 2's have been divided by 7.0. We normalize so we can compare the cost per unit traded.  
    \item Each row within a block displays normalized costs for a given \textit{evaluation kappa} computed using the cost functions $C_1$ and $C_2$, see equations \eqref{eq:actual-cost-1}-\eqref{eq:actual-cost-2}. In the table the final four columns display the costs for firm 1 and 2, the first of which is the \textit{actual cost}, using the evaluation $\kappa$ for its row. 
    \item \textit{Example:} The fourth column from the right in the table represents $C_1(s_1^k, s_2^m; \kappa)$ when firm 1's active type is $k$ and firm 2's is $m$ and the evaluation  $\kappa$ is shown in the column labeled "Eval $\kappa$". 
    \item The first row represents active types for firm 1 and 2 of 1 and 1 and an evaluation $\kappa$ of 1.0. The \textit{actual cost} for firm 1 using $\kappa=1$ is 15.81. On the other hand, the second column from the right shows $C_2(s_2^m, s_1^l;k)$ and a normalized cost of 14.68. Thus firm 1's cost per unit traded is modestly more than that of firm 2. 
    \item The columns labeled "Exp." in the third column from the right and the last column of the table are the expected costs to each strategy, $E_1(s_1^k, s_2^*; \kappa_1^k)$ and $E_2(s_2^m, s_1^*; \kappa_2^m)$ in equations \eqref{eq:exp-cost1}-\eqref{eq:exp-cost2}. Recall that these are the basis upon which the strategies were formed. In the cast of firms 1 and 2 having active types 1 and 1, the expected costs are 14.71 and 21.88 respectively. 
\end{enumerate}

\renewcommand{\arraystretch}{1.25}
\begin{table}[h!]
\centering
\begin{tabular}{|c|c|c|c|c|c|c|c|c|c|c|}
\hline
\multicolumn{3}{|c|}{\text{Firm 1}} & \multicolumn{3}{c|}{\text{Firm 2}} &  & \multicolumn{2}{c|}{\text{Firm 1}} & \multicolumn{2}{c|}{\text{Firm 2}} \\
\text{Type} & \text{$\kappa$} & \text{$f$} & \text{Type} & \text{$\kappa$} & \text{$f$} & \ Eval $\kappa_e$ & \text{Cost} & \text{Exp.}  & \text{Cost} & \text{Exp.} \\
\hline
$t_1^1$ & 1.00 & 3.00 & $t_2^1$ & 2.00 & 7.00 & 1.00 & 15.81 & 14.71  & 14.68 & 21.88 \\
 & 1.00 & 3.00 &  & 2.00 & 7.00 & 2.00 & 21.73 &   & 19.28 &  \\
 & 1.00 & 3.00 &  & 2.00 & 7.00 & 3.00 & 27.65 &   & 23.89 &  \\
 & 1.00 & 3.00 &  & 2.00 & 7.00 & 15.00 & 98.68 &   & 79.16 &  \\
\hline$t_1^1$ & 1.00 & 3.00 & $t_2^2$ & 15.00 & 5.00 & 1.00 & 12.49 & 14.71  & 22.64 & 57.10 \\
 & 1.00 & 3.00 &  & 15.00 & 5.00 & 2.00 & 19.12 &   & 25.05 &  \\
 & 1.00 & 3.00 &  & 15.00 & 5.00 & 3.00 & 25.76 &   & 27.47 &  \\
 & 1.00 & 3.00 &  & 15.00 & 5.00 & 15.00 & 105.37 &   & 56.49 &  \\
\hline$t_1^2$ & 3.00 & 5.00 & $t_2^1$ & 2.00 & 7.00 & 1.00 & 18.62 & 31.34  & 17.85 & 21.88 \\
 & 3.00 & 5.00 &  & 2.00 & 7.00 & 2.00 & 25.02 &   & 23.56 &  \\
 & 3.00 & 5.00 &  & 2.00 & 7.00 & 3.00 & 31.42 &   & 29.28 &  \\
 & 3.00 & 5.00 &  & 2.00 & 7.00 & 15.00 & 108.22 &   & 97.85 &  \\
\hline$t_1^2$ & 3.00 & 5.00 & $t_2^2$ & 15.00 & 5.00 & 1.00 & 16.98 & 31.34  & 25.98 & 57.10 \\
 & 3.00 & 5.00 &  & 15.00 & 5.00 & 2.00 & 24.14 &   & 28.83 &  \\
 & 3.00 & 5.00 &  & 15.00 & 5.00 & 3.00 & 31.29 &   & 31.67 &  \\
 & 3.00 & 5.00 &  & 15.00 & 5.00 & 15.00 & 117.14 &   & 65.80 &  \\
\hline\end{tabular}
\caption{Costs for firms 1 and 2 for the case of two firms and 2 types. The costs displayed have been normalized.}
\label{tab:two-firms-2-types-1-normalized}
\end{table}

\renewcommand{\arraystretch}{1.25}
\begin{table}[h!]
\centering
\begin{tabular}{|c|c|c|c|c|c|c|c|c|c|c|}
\hline
\multicolumn{3}{|c|}{\text{Firm 1}} & \multicolumn{3}{c|}{\text{Firm 2}} &  & \multicolumn{2}{c|}{\text{Firm 1}} & \multicolumn{2}{c|}{\text{Firm 2}} \\
\text{Type} & \text{$\kappa$} & \text{$f$} & \text{Type} & \text{$\kappa$} & \text{$f$} & \ Eval $\kappa_e$ & \text{Cost} & \text{Exp.}  & \text{Cost} & \text{Exp.} \\
\hline
$t_1^1$ & 1.0 & 3.0 & $t_2^1$ & 2.0 & 7.0 & 1.0 & 47.4 & 44.1  & 102.7 & 153.2 \\
 & 1.0 & 3.0 &  & 2.0 & 7.0 & 2.0 & 65.2 &   & 135.0 &  \\
 & 1.0 & 3.0 &  & 2.0 & 7.0 & 3.0 & 83.0 &   & 167.2 &  \\
 & 1.0 & 3.0 &  & 2.0 & 7.0 & 15.0 & 296.0 &   & 554.1 &  \\
\hline$t_1^1$ & 1.0 & 3.0 & $t_2^2$ & 15.0 & 5.0 & 1.0 & 37.5 & 44.1  & 113.2 & 285.5 \\
 & 1.0 & 3.0 &  & 15.0 & 5.0 & 2.0 & 57.4 &   & 125.3 &  \\
 & 1.0 & 3.0 &  & 15.0 & 5.0 & 3.0 & 77.3 &   & 137.4 &  \\
 & 1.0 & 3.0 &  & 15.0 & 5.0 & 15.0 & 316.1 &   & 282.4 &  \\
\hline$t_1^2$ & 3.0 & 5.0 & $t_2^1$ & 2.0 & 7.0 & 1.0 & 93.1 & 156.7  & 124.9 & 153.2 \\
 & 3.0 & 5.0 &  & 2.0 & 7.0 & 2.0 & 125.1 &   & 164.9 &  \\
 & 3.0 & 5.0 &  & 2.0 & 7.0 & 3.0 & 157.1 &   & 204.9 &  \\
 & 3.0 & 5.0 &  & 2.0 & 7.0 & 15.0 & 541.1 &   & 684.9 &  \\
\hline$t_1^2$ & 3.0 & 5.0 & $t_2^2$ & 15.0 & 5.0 & 1.0 & 84.9 & 156.7  & 129.9 & 285.5 \\
 & 3.0 & 5.0 &  & 15.0 & 5.0 & 2.0 & 120.7 &   & 144.1 &  \\
 & 3.0 & 5.0 &  & 15.0 & 5.0 & 3.0 & 156.4 &   & 158.3 &  \\
 & 3.0 & 5.0 &  & 15.0 & 5.0 & 15.0 & 585.7 &   & 329.0 &  \\
\hline\end{tabular}
\caption{Costs for Firms 1 and 2 for the case of two firms and 2 types. The costs displayed have not been normalized.}
\label{tab:two-firms-2-types-1}
\end{table}

We repeat the Table \ref{tab:two-firms-2-types-1-normalized} but  this time with the costs \textit{not normalized}. We now discuss implementation costs of the strategies in a little more detail. Consider when firm 1's active type is 1. Then its normalized  \textit{expected} cost is 14.71 and in this case its believes that $\kappa=1.0$, meaning this is the value of the market impact parameter used in selecting its type 1 strategy. Meanwhile its \textit{actual} cost depends on a few things, first among them is what the active type firm 2 is, which firm 1 does have ex ante knowledge of. In addition, it depends on the actual $\kappa$ that determines how the firms' trading impacts market prices\footnote{More generally the actual costs depend on the actual microstructure details governing how the market absorbs trades, and as discussed earlier one can make a more expansive Bayesian game that captures this uncertainty as well.}. 

\smallsec{Cumulative cost analysis}

\begin{figure}[h!]
    \centering
    \includegraphics[width=1.0\linewidth]{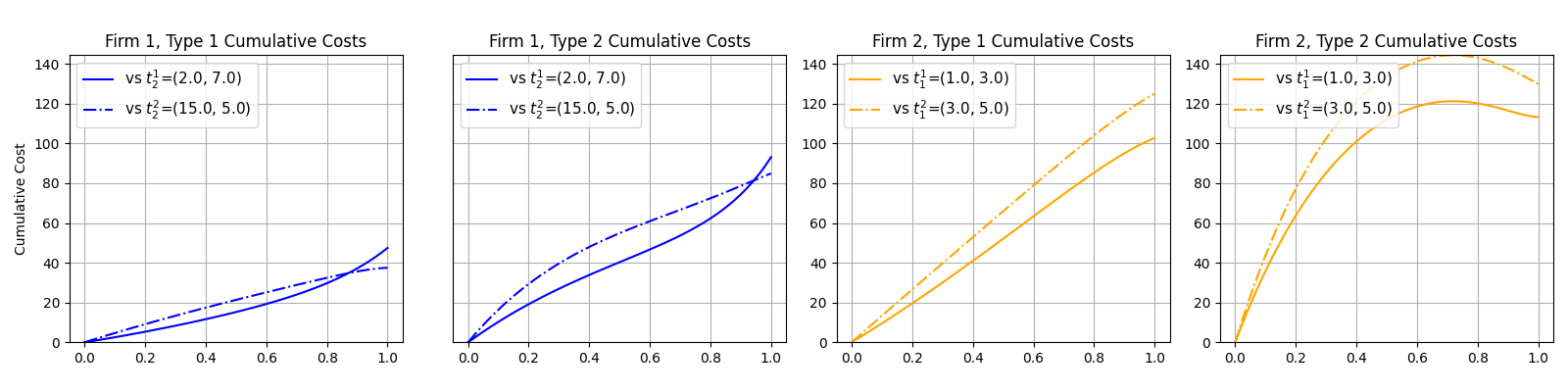}
    \caption{Cumulative costs for the trading strategies depicted in Figure \ref{fig:two-firms-2-types-1} with total implementation costs as in Table \ref{tab:two-firms-2-types-1} for the evaluation $\kappa$ set to 1.0. One can compare these costs to the implementation costs in Table \ref{tab:two-firms-2-types-1}.} 
    \label{fig:two-firms-2-types-cum-1}
\end{figure}

We now look at the cumulative costs for each firm and active type  of the strategies in Figure \ref{fig:two-firms-2-types-1} (see Section \ref{sec:cumulative-costs} regarding cumulative costs). We plot the cumulative costs in Figure \ref{fig:two-firms-2-types-cum-1} wherein each plot shows the cumulative costs for a given firm and active type and then its cumulative costs versus both possible active types of the other firm. In the figure an evaluation $\kappa$ of 1 was used, corresponding to the top row of each block in Table \ref{tab:two-firms-2-types-1}.

Now referencing Table \ref{tab:two-firms-2-types-1}, we see that when firm 1's active type is 1 then firm 1's implementation costs are 15.81 and 14.68  when firm 2's active type is 1 or 2 respectively. At first blush this seems odd because consulting Figure \ref{fig:two-firms-2-types-1}, when firm 2's active type is 2 it trades considerably more aggressively up to time $t=0.6$ (that is, through 60\% of the total time trading). Should that not push prices up so that when firm 1 as type 1 faces firm 2 as type 2 it pays more than when firm 2 is type 1? 

Though the confusion is understandable, its resolution arises from that facts that (i) firm 2 also sells more rapidly from $t=0.6$ to $t=1.0$ when it is type 2 versus type 1, and (ii) that the evaluation $\kappa$ is set relatively low at 1.0. To see the impact of the first point, we consult the top-left plot in Figure \ref{fig:two-firms-2-types-cum-1}. In that plot, the solid blue line represents firm 1's cumulative costs versus firm 2 when it is type 1, while the dash-dotted blue line is when firm 2 is type 2. Notice that the latter \textit{is} more costly until $t$ is approximately 0.8 after which the cumulative costs flip. A careful look as that the costs versus firm 2 as type 2 rate of increase modestly \textit{moderates} starting at $t\approx 0.8$ while the costs facing firm 2 as type 1 start to increase more rapidly. 

We can see why by looking at the trading trajectories in Figure \ref{fig:two-firms-2-types-1}. When firm 2 is type 1 it continues to buy stock from $t=0.8$ to $t=1.0$ exactly when firm 1 (as type 1) begins to trade more rapidly. Meanwhile starting at a time well before $t=0.8$ when firm 2 is type 2, it begins to sell and its pace of selling becomes quite rapid starting at $t=0.8$. It is the effect of this rapid selling that causes the moderation in price increases we see in the dash-dotted line in the top-left plot of Figure \ref{fig:two-firms-2-types-1}.

\vspace{-15pt}

\begin{figure}[h!]
    \centering
    \includegraphics[width=1.0\linewidth]{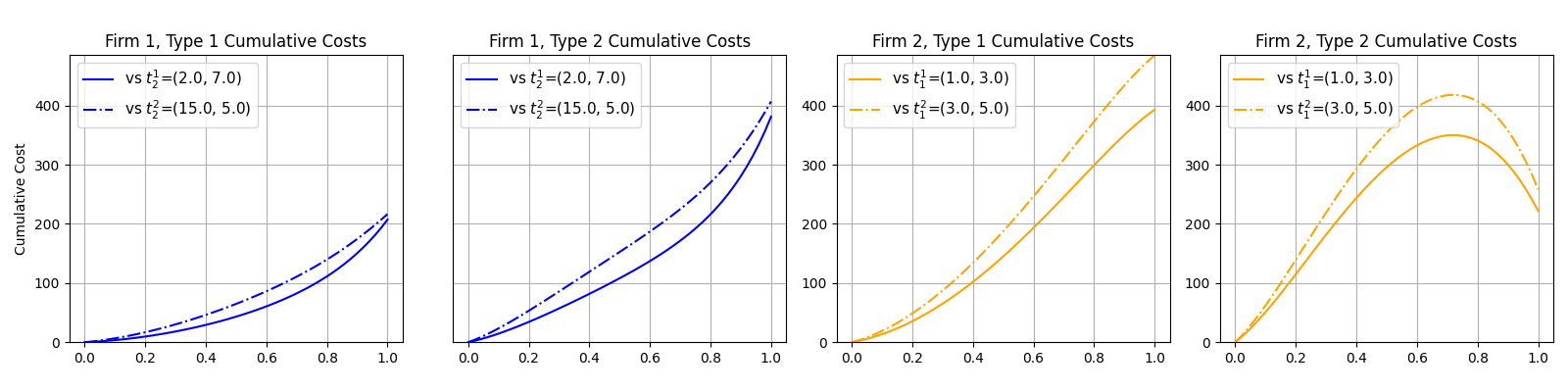}
    \caption{Analogous to Figure \ref{fig:two-firms-2-types-cum-1} but with $\kappa=10.0$. This plot shows the cumulative costs for the trading strategies depicted in Figure \ref{fig:two-firms-2-types-1} with total implementation costs as in Table \ref{tab:two-firms-2-types-1} for the evaluation $\kappa$ set to 10.0.}
    \label{fig:two-firms-2-types-cum-kappa-10}
\end{figure}

We repeat the cumulative costs plots in Figure \ref{fig:two-firms-2-types-cum-kappa-10} but with an evaluation $\kappa$ set to 10.0

\subsection{Two firms and three types}
\label{sec:ex-two-firms-three-types-1}

This section largely repeats the approach of Section \ref{sec:two-firm-examples-1} and we very briefly examine the case of two firms with three possible types for each firm. In Table \ref{tab:two-firms-3-types-1} below displays the type data for this scenario. 

\medskip

\begin{table}[h!]
\centering
\renewcommand{\arraystretch}{1.33} 
\begin{tabular}{llll}
\hline
 & \multicolumn{1}{c}{\text{Type 1}} & \multicolumn{1}{c}{\text{Type 2}} & \multicolumn{1}{c}{\text{Type 3}} \\
\hline
\text{firm 1} & $t_1^{1}=(20.0, 1.0)$ & $t_1^{2}=(5.0, 2.0)$ & $t_1^{3}=(1.0, 3.0)$ \\
\text{firm 2} & $t_2^{1}=(5.0, 0.2)$ & $t_2^{2}=(2.0, 1.5)$ & $t_2^{3}=(1.0, 5.0)$ \\
\hline
\end{tabular}
\caption{Type data for two firms. Each type $t_i^k=(\kappa_i^k, f_i^k)$ consists of a market impact parameter $\kappa_i^k$ and a target size $f_i^k$. For example, firm 1's type 1 has $\kappa_1^1=20.0$ and $f_1^1=1.0$.}
\label{tab:two-firms-3-types-1}
\end{table}

\medskip

 Equation \eqref{eq:two-firms-3-types-1} displays the common prior matrix $P$ (see Section \ref{sec:types}) and the conditional probability matrices, see equations \eqref{eq:cond-prob-1} and \eqref{eq:cond-prob-2}, $P_1$ and $P_2$.

\begin{equation}
\label{eq:two-firms-3-types-1}
P = \begin{pmatrix}
0.15 & 0.10 & 0.10 \\
0.15 & 0.20 & 0.10 \\
0.05 & 0.05 & 0.10
\end{pmatrix} \!,\quad P_1 = \begin{pmatrix}
0.43 & 0.29 & 0.29 \\
0.33 & 0.44 & 0.22 \\
0.25 & 0.25 & 0.50
\end{pmatrix} \!,\quad P_2 = \begin{pmatrix}
0.43 & 0.43 & 0.14 \\
0.29 & 0.57 & 0.14 \\
0.33 & 0.33 & 0.33
\end{pmatrix}
\end{equation} 

\begin{figure}[h!]
    \centering
    \includegraphics[width=0.75\linewidth]{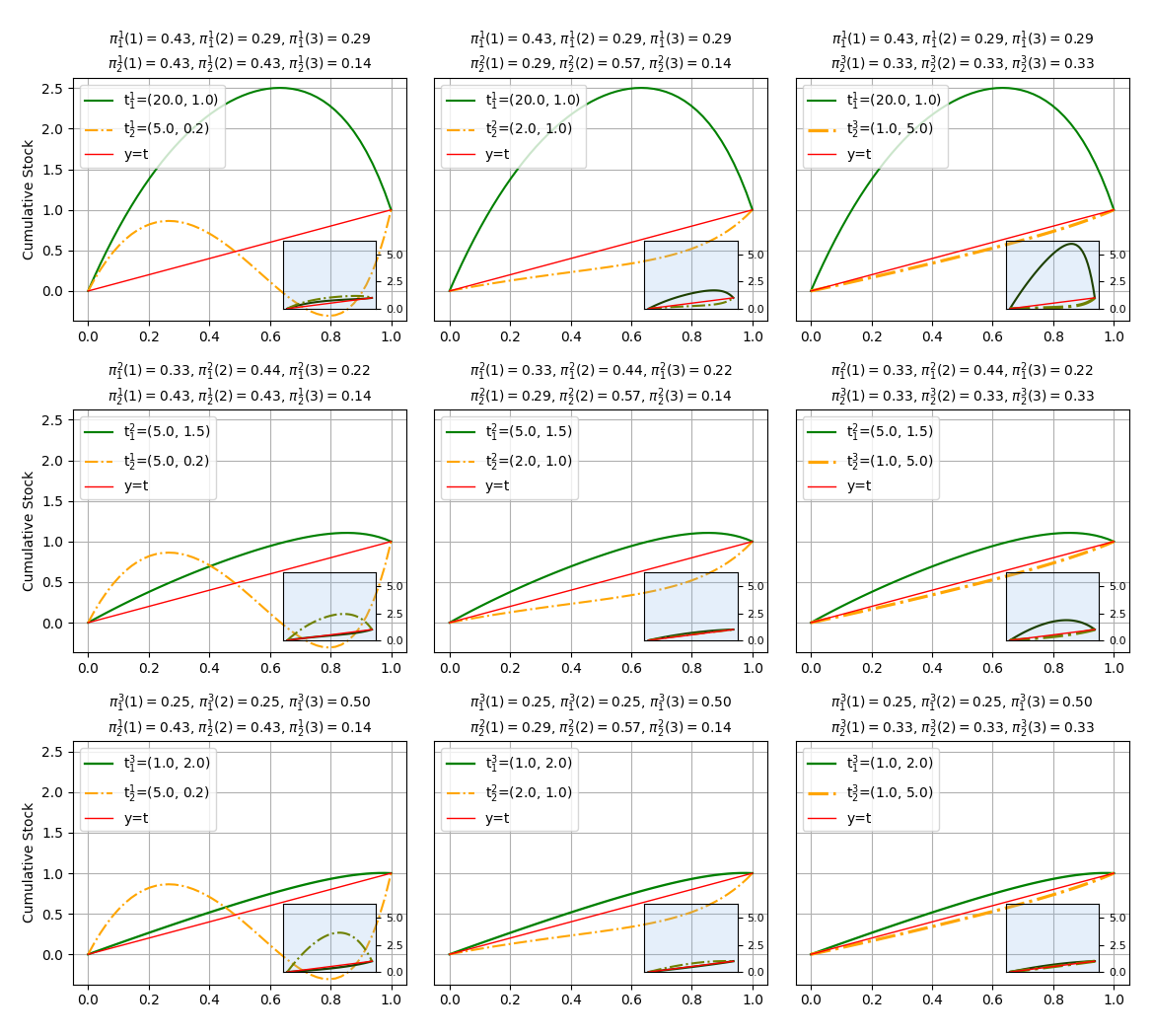}
    \caption{Strategy pairs for two firms with two possible types each using the type data shown in Table \ref{tab:two-firms-3-types-1} and probability matrices in \cref{eq:two-firms-3-types-1}. Each sub-plot shows the trading strategies of firms 1 and 2 when their active types are $k$ and $m$ according to the sub-plot's location. For example, the top-left plot shows the strategies when the active types are both 1. The inset in each plot shows the strategy pair for firms 1 and 2 when there is only one possible type and the type data is the same as the active types in the primary plot. In each plot, solid lines are firm 1's strategy and dash-dotted lines are firm 2's.}
    \label{fig:two-firms-3-types-1}
\end{figure}

Figure \ref{fig:two-firms-3-types-1} displays the trading strategies for firms 1 and 2 in a 3x3 plot grid, each sub-plot showing one possible pair of active types for the firms, with the $(i,j)$ subplot showing the case where firm 1's active type is $i$ and firm 2's is $j$.

Given the strategies above, we discuss their implementation costs as described in Section \ref{sec:performance-evaluation}. In Table \ref{tab:two-firms-3-types-1} we display the costs and expected costs as follows.

\renewcommand{\arraystretch}{1.15}
\begin{table}[h!]
\centering
\begin{tabular}{|c|c|c|c|c|c|c|c|c|c|c|}
\hline
\multicolumn{3}{|c|}{\text{firm 1}} & \multicolumn{3}{c|}{\text{firm 2}} &  & \multicolumn{2}{c|}{\text{firm 1}} & \multicolumn{2}{c|}{\text{firm 2}} \\
\text{Type} & \text{$\kappa$} & \text{$f$} & \text{Type} & \text{$\kappa$} & \text{$f$} & \ Eval $\kappa_e$ & \text{Cost} & \text{Exp.}  & \text{Cost} & \text{Exp.} \\
\hline
$t_1^1$ & 20.00 & 1.00 & $t_2^1$ & 5.00 & 0.25 & 1.00 & 26.73 & 4.85  & 2.09 & 6.27 \\
 & 20.00 & 1.00 &  & 5.00 & 0.25 & 2.00 & 27.64 &   & 1.57 &  \\
 & 20.00 & 1.00 &  & 5.00 & 0.25 & 5.00 & 30.37 &   & 0.01 &  \\
 & 20.00 & 1.00 &  & 5.00 & 0.25 & 20.00 & 44.03 &   & -7.80 &  \\
\hline$t_1^1$ & 20.00 & 1.00 & $t_2^2$ & 2.00 & 1.50 & 1.00 & 24.01 & 4.85  & 3.40 & 7.87 \\
 & 20.00 & 1.00 &  & 2.00 & 1.50 & 2.00 & 23.21 &   & 6.01 &  \\
 & 20.00 & 1.00 &  & 2.00 & 1.50 & 5.00 & 20.83 &   & 13.84 &  \\
 & 20.00 & 1.00 &  & 2.00 & 1.50 & 20.00 & 8.91 &   & 53.03 &  \\
\hline$t_1^1$ & 20.00 & 1.00 & $t_2^3$ & 1.00 & 5.00 & 1.00 & 22.67 & 4.85  & 9.71 & 10.01 \\
 & 20.00 & 1.00 &  & 1.00 & 5.00 & 2.00 & 18.43 &   & 14.16 &  \\
 & 20.00 & 1.00 &  & 1.00 & 5.00 & 5.00 & 5.70 &   & 27.51 &  \\
 & 20.00 & 1.00 &  & 1.00 & 5.00 & 20.00 & -57.97 &   & 94.23 &  \\
\hline$t_1^2$ & 5.00 & 2.00 & $t_2^1$ & 5.00 & 0.25 & 1.00 & 4.18 & 11.87  & 5.95 & 6.27 \\
 & 5.00 & 2.00 &  & 5.00 & 0.25 & 2.00 & 5.34 &   & 6.76 &  \\
 & 5.00 & 2.00 &  & 5.00 & 0.25 & 5.00 & 8.83 &   & 9.18 &  \\
 & 5.00 & 2.00 &  & 5.00 & 0.25 & 20.00 & 26.30 &   & 21.29 &  \\
\hline$t_1^2$ & 5.00 & 2.00 & $t_2^2$ & 2.00 & 1.50 & 1.00 & 5.35 & 11.87  & 5.54 & 7.87 \\
 & 5.00 & 2.00 &  & 2.00 & 1.50 & 2.00 & 6.68 &   & 7.85 &  \\
 & 5.00 & 2.00 &  & 2.00 & 1.50 & 5.00 & 10.67 &   & 14.78 &  \\
 & 5.00 & 2.00 &  & 2.00 & 1.50 & 20.00 & 30.62 &   & 49.44 &  \\
\hline$t_1^2$ & 5.00 & 2.00 & $t_2^3$ & 1.00 & 5.00 & 1.00 & 9.65 & 11.87  & 10.89 & 10.01 \\
 & 5.00 & 2.00 &  & 1.00 & 5.00 & 2.00 & 11.95 &   & 14.87 &  \\
 & 5.00 & 2.00 &  & 1.00 & 5.00 & 5.00 & 18.83 &   & 26.82 &  \\
 & 5.00 & 2.00 &  & 1.00 & 5.00 & 20.00 & 53.27 &   & 86.54 &  \\
\hline$t_1^3$ & 1.00 & 3.00 & $t_2^1$ & 5.00 & 0.25 & 1.00 & 4.95 & 8.56  & 8.18 & 6.27 \\
 & 1.00 & 3.00 &  & 5.00 & 0.25 & 2.00 & 6.59 &   & 9.64 &  \\
 & 1.00 & 3.00 &  & 5.00 & 0.25 & 5.00 & 11.51 &   & 14.00 &  \\
 & 1.00 & 3.00 &  & 5.00 & 0.25 & 20.00 & 36.10 &   & 35.82 &  \\
\hline$t_1^3$ & 1.00 & 3.00 & $t_2^2$ & 2.00 & 1.50 & 1.00 & 6.51 & 8.56  & 7.27 & 7.87 \\
 & 1.00 & 3.00 &  & 2.00 & 1.50 & 2.00 & 8.51 &   & 10.01 &  \\
 & 1.00 & 3.00 &  & 2.00 & 1.50 & 5.00 & 14.53 &   & 18.24 &  \\
 & 1.00 & 3.00 &  & 2.00 & 1.50 & 20.00 & 44.59 &   & 59.36 &  \\
\hline$t_1^3$ & 1.00 & 3.00 & $t_2^3$ & 1.00 & 5.00 & 1.00 & 11.39 & 8.56  & 12.37 & 10.01 \\
 & 1.00 & 3.00 &  & 1.00 & 5.00 & 2.00 & 14.82 &   & 16.71 &  \\
 & 1.00 & 3.00 &  & 1.00 & 5.00 & 5.00 & 25.13 &   & 29.73 &  \\
 & 1.00 & 3.00 &  & 1.00 & 5.00 & 20.00 & 76.68 &   & 94.80 &  \\
\hline\end{tabular}
\caption{Costs for firms 1 and 2 for the case of two firms and 3 types. The costs displayed have been normalized.}
\label{tab:two-firms-3-types-1}
\end{table}

\subsection{Two firms trading in the presence of a non-strategic firm}
\label{sec:non-strategic-example}

In this section we provide several illustrations of when there are two firms in competition both of who weigh the possibility of a third, non-strategic firm, the details of which were introduced in Section \ref{sec:non-strategic-firm}. To begin we introduce the type data and unlike the previous two-firm examples we include a third parameter for each type which is the target size each firm and type believes the non-strategic firm will trade. In these examples, the non-strategic firm is assumed to always trade a risk-neutral strategy, the strategy that maximally spaces trades out over time. 

Table \ref{tab:two-firms-2-types-goliath-1} displays the type information for our first scenario. This example is as simple as can be. Each firm has two types and the types are virtually identical. All firms believe the market impact parameter $\kappa$ is equal to 1.0 and all firms trade to a target size of 1.0. The \textit{only} difference among types is that firm 2's type 1 believes there is a non-strategic trader trading to a target size of 5.0 (as a reference this is the third element of $t_2^1$ in Table \ref{tab:two-firms-2-types-goliath-1}) below). We show below that even this difference profoundly impacts the strategies each type trades.

\begin{table}[h!]
\centering
\renewcommand{\arraystretch}{1.25} 
\begin{tabular}{lll}
\hline
 & \multicolumn{1}{c}{\text{Type 1}} & \multicolumn{1}{c}{\text{Type 2}} \\
\hline
\text{Firm 1} & $t_1^{1}=(1.0, 1.0, 0.0)$ & $t_1^{2}=(1.0, 1.0, 0.0)$ \\
\text{Firm 2} & $t_2^{1}=(3.0, 1.0, 5.0)$ & $t_2^{2}=(1.0, 1.0, 0.0)$ \\
\hline
\end{tabular}
\caption{Type data for firms 1 and 2. Each type $t_i^k=(\kappa_i^k, f_i^k, b_1^k)$ consists of a market impact parameter $\kappa_i^k$, target size $f_i^k$ and the size of the non-strategic firm. For example, firm 1's type 1 has $\kappa_1^1=1.0$, $f_1^1=1.0$ and $b_1^1=0.0$, the latter meaning that firm 1's type 1 does not believe the non-strategic firm will "show up".}
\label{tab:two-firms-2-types-goliath-1}
\end{table}

\begin{figure}[h!]
    \centering
    \includegraphics[width=1\linewidth]{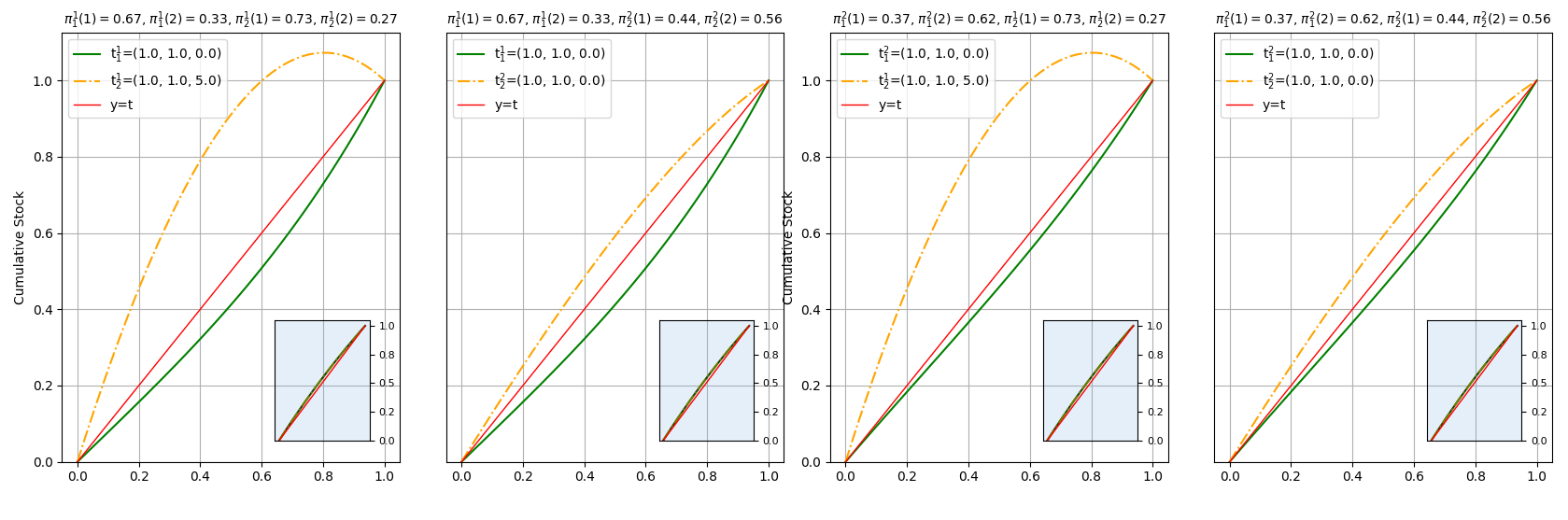}
    \caption{Two firms in competition when there is \textit{possibly} a third non-strategic firm. In this set of plots both firms have two possible types and all types have the same market impact parameter and same target quantity. The type data for this scenario is displayed in the plots and also in Table \ref{tab:two-firms-2-types-goliath-1}. The only difference among all types is that firm 2, Type 1 believes that there is a non-strategic investor trading with a target quantity of 5.0. As a result firm 2's type 1 trades an eager, over-buying strategy and both of firm 1's types trade a slightly risk-averse strategy, the differences owing to the conditional probabilities of which firm 2 active type will show up. Note that the knock-on effect that firm 2's type 2 which does not believe there is a non-strategic trade, still trades a modestly eager strategy.}
    \label{fig:goliath-1}
\end{figure}

Figure \ref{fig:goliath-1} is displays how each firm trades given the type data in Table \ref{tab:two-firms-2-types-goliath-1}. The inset of each plot shows how the firms would trade without any consideration of the non-strategic firm in which case the firms trade more or less a straight-line risk-neutral strategy, though very slightly eagerly. The main takeaway is that the inclusion of a potential non-strategic firm impacts \textit{all} of the firms' trading. Because firm 2's type 1 believes there is a non-strategic trader, it trades eagerly to get ahead of this trader. But firm 1's type 1 places a 67\% probability that firm 2 is represented by its type 1 and therefore trades a risk-averse strategy. Firm 2's type 2 which does not believe there is a non-strategic firm nevertheless trades an eager strategy (though less so than firm 2's type 1) because of firm 1's trading strategy.

\vspace{-10pt}

\begin{figure}[h!]
    \centering
    \includegraphics[width=1\linewidth]{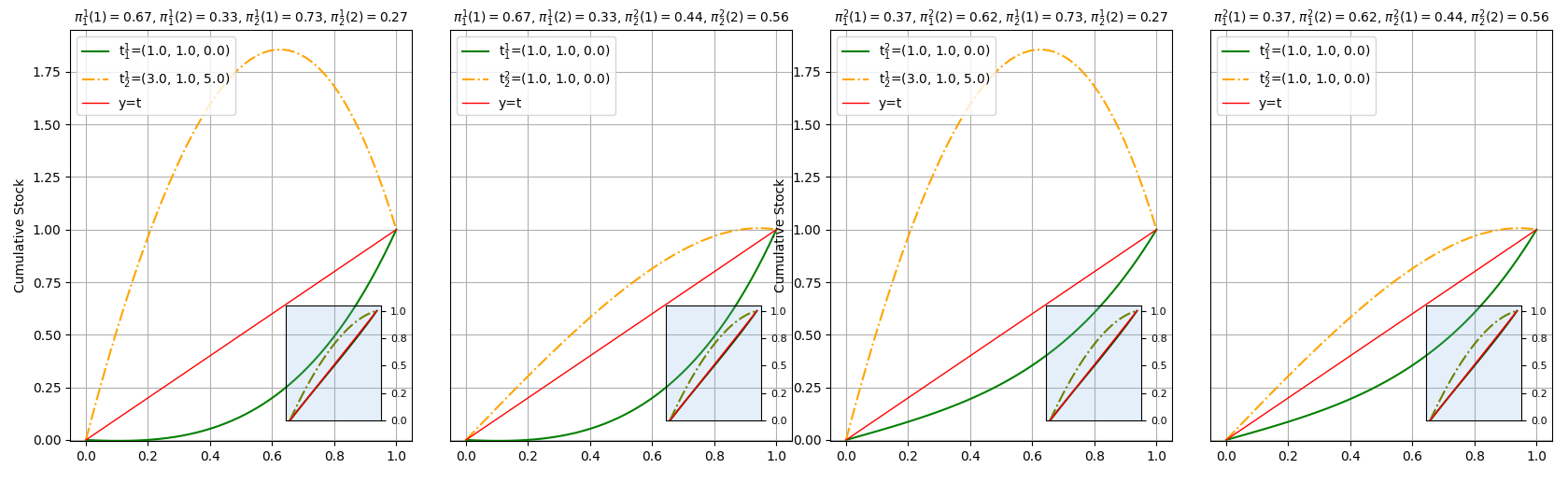}
    \caption{Two firms almost identical to Figure \ref{fig:goliath-1} except that firm 2's type 1 now uses a market impact parameter of 3.0 as opposed to 1.0, meaning that the non-strategic firm's impact on the market is significantly greater. Not surprisingly firm 2's type 1 trades significantly more eagerly and overbuys significantly more.}
    \label{fig:goliath-2}
\end{figure}

In the next plots, Figure \ref{fig:goliath-2}, we repeat the scenario in the previous Figure \ref{fig:goliath-2}, but this time change firm 2's type 1's market impact parameter from 1.0 to 3.0. The upshot of this is that 

\appendix

\section{Mathematical reference}

Some standard facts from basic linear algebra and differential equations are used throughout so we reference them here.

\subsection{Trading strategies and linear ordinary differential equations}
\label{ap:odes}

This appendix contains basic facts about linear systems of ordinary differential equations used within the paper. We will be concerned \textit{primarily} with functions of the form:

\begin{equation}
    s = \pmat{s_1 \\ \vdots \\ s_N}, \qquad s_i: [0, 1] \to \R
\end{equation}

Thus $s$ represents a vector of \textit{independent} (that is, non-interacting) functions, each from the unit interval $[0,1]$ to the real numbers. In the applications in this paper, a function $s_i$ typically represents a trading strategy, namely a description of the cumulative purchase or sale from time $0$ to some time $t \in [0, 1]$. Our concern will often by to study the first and second derivatives of $s$. To facilitate this we adopt a standard notation $v = \dot s$ and write:

\begin{equation}
    v = \pmat{v_1\\ \vdots \\ v_n}, \quad \dot v = \pmat{\dot v_1 \\ \vdots \\ \dot v_n}
\end{equation}

The typical scenario in this paper is that there is a system of differential equations involving $s$ that can be transformed to one involving a \textit{non-singular} matrix $M$ involving $v$ as follows:

\begin{equation}
    \dot v = M v 
\end{equation}

Generally we know boundary conditions on $s$ and these take the form $s(0)=0$ and $s(1)=f$, where $f$ is a vector $f=(c_f\, \cdots \, c_f)^\top$. The standard approach to finding $s$ is to first find $v$ in terms of an initial condition for $v$, $v_0 := v(0)$ and then find $v_0$ in terms of $f$, which we explain now. To begin, note that the following is \textit{always} true:

\begin{equation}
    \label{eq:first-order-soln}
    v(t) = e^{Mt} v_0 \qquad (\text{Note: $v(t)=\dot s(t)$})
\end{equation}

In order to compute $s(t) = \int v(u)\du$ we use the fact that $\int_0^t e^{Mu} \du = (e^{Mt} - I)M^{-1}$ holds whenever $M$ is non-singular. Therefore:

\begin{equation}
    s(t) = \int_0^t e^{Mu} v_0 \du = (e^{Mt} - I) M^{-1}v_0
\end{equation}

From this we may compute $v_0$ in terms of $f$ as:

\begin{equation}
\label{eq:v_0}
    v_0 =  M (e^{M} - I)^{-1} f
\end{equation}

and therefore the final expression for $s(t)$ may be written simply as:

\begin{equation}
    \label{eq:s-expr-appendix}
    s(t) = (e^{Mt} - I) M^{-1} v_0
\end{equation}

Finally we note that at times we will use the following two identities for a square matrix $M$:

\begin{equation}
    M e^M = e^M M, \qquad M^{-1} e^M = e^M M^{-1}
\end{equation}

and therefore

\begin{equation}
    (e^{Mt} - I) M^{-1} = M^{-1}(e^{Mt} - I)
\end{equation}

\section{Acknowledgments}

    I extend my sincere thanks to the Machine Learning Research Group at Morgan Stanley who both originally suggested exploring Bayesian games and hosted my seminar on the topic during which they made quite a few important suggestions. For version two, a special thanks to Yuriy Nevmyvaka of Morgan Stanley's for a careful reading of the first draft and identifying numerous typos which were corrected in the second draft.

\bibliographystyle{alpha}
\bibliography{bayesian-position-building}

\end{document}